\documentclass[pra, showpacs, twocolumn, floatfix]{revtex4}
\usepackage{epsfig}

\begin{document}

\title{Muon pair creation from positronium in a linearly polarized laser field}

\author{Carsten M\"uller, Karen Z. Hatsagortsyan, and Christoph H. Keitel}
\affiliation{ Max-Planck-Institut f\"ur Kernphysik, 
Saupfercheckweg 1, D-69117 Heidelberg, Germany}

\date{\today}

\begin{abstract}
Positronium decay into a muon-antimuon pair by virtue of the interaction with a superintense laser field of linear polarization is considered. The minimum laser intensity required amounts to a few 10$^{22}$\,W/cm$^2$ in the near-infrared frequency range. Within the framework of laser-dressed quantum electrodynamics, the total reaction rate is calculated and related to the cross section for field-free electron-positron annihilation into muons. The muons are created with ultrarelativistic energies and emitted under narrow angles along the laser propagation direction. The dynamical properties of the muons are interpreted in terms of a classical simple man's model for the production process. We show that the most promising setup for an experimental investigation of the process employs two counterpropagating laser beams impinging on a positronium target, where the advantage of the coherent electron-positron collisions becomes evident.
\end{abstract}


\maketitle


\section{Introduction}
Inside the most powerful laser fields available today \cite{Bahk} electrons and positrons can acquire kinetic energies in the GeV range. Next-generation laser devices are  envisaged to generate even higher electromagnetic field intensities of $\sim10^{23}$\,W/cm$^2$ and above \cite{ELI}. The GeV energy domain has also been demonstrated accessible via laser-wakefield acceleration of electrons \cite{accel} and positrons \cite{pos}. In principle, such high energies can be exploited to induce elementary particle reactions like heavy lepton-pair or hadron production in $e^+e^-$ collisions. The energy scale associated with superintense laser fields is thus more typical for particle physics than for atomic and molecular physics, where laser fields are employed traditionally. Apart from the high energies achievable, lasers can be utilized to generate well-controlled particle collisions at microscopically small impact parameters, which can lead to high luminosities \cite{Corkum,Guido,collider}. Ions (protons) can also be accelerated by lasers \cite{ion_accel}, which may be used for triggering hadronic processes, such as pion production \cite{pion}. Consequently, intense laser fields offer alternative and supplementary ways towards high-energy physics \cite{collider,report,help2,muonPRD,muonPLB}. Similar efforts to merge laser physics with nuclear physics are being undertaken successfully \cite{Schwoerer}. Laser-induced nuclear fusion \cite{fusion}, photofission \cite{fission}, and neutron production \cite{nuc} have already been observed in experiment, whereas other processes like coherent nuclear excitation \cite{Thomas} or muon-catalyzed fusion with laser assistance \cite{muonfusion} have been discussed on theoretical grounds.

One of the most elementary processes in particle physics is the creation of a muon-antimuon pair from electron-positron annihilation, which has proven fundamental for the understanding of other $e^+e^-$ reactions \cite{Peskin}. The threshold particle energy in the center-of-mass (c.m.) frame required for this reaction is determined by the muon rest mass $M\approx 100$\,MeV/$c^2$ and amounts to $E_{\rm cm}=2Mc^2$. In conventional accelerator facilities the collision energy is provided by quasistatic electric and guiding magnetic fields.

Motivated by the sustained progress in laser technology, we have addressed the question whether the process $e^+e^-\to\mu^+\mu^-$ is feasible with low-energy electrons and positrons in the presence of an intense laser field \cite{muonPRD,muonPLB}. The electrons and positrons are assumed to form initially a nonrelativistic $e^+e^-$ plasma or a gas of positronium (Ps) atoms \cite{stable}. Since the initial energy of the particles is far below $2Mc^2$, their annihilation into a muon pair cannot happen without the influence of the external field. We stress, however, that the muon pair is not produced by the laser field itself, as it has been achieved before with regard to $e^+e^-$ pair creation in the collision of a laser and an electron beam \cite{Ritus, SLAC}. In our case, this would require unrealistically strong laser fields of the order of the critical field $E_{\rm cr}^{(\mu)}=M^2c^3/e\hbar$ for the muon, with the elementary charge $e$ and Planck's constant $\hbar$, which corresponds to laser intensities $I\sim 10^{38}$\,W/cm$^2$. Instead, the muon pair is produced in an annihilating electron-positron collision, where the role of the laser field is to supply the required energy to the colliding particles.

The case with initial Ps atoms is particularly interesting since it allows for {\it coherent} $e^+e^-$ collisions, which are characterized by microscopically small impact parameters \cite{collider}. This is due to the fact that the electron and positron are initially confined to the atomic size of the bound Ps state and then, after instantaneous ionization, are coherently accelerated by the strong laser field, which drives the particles into opposite directions along the laser electric-field component. This leads to periodic $e^+e^-$ (re)collisions whose energy is proportional to the laser intensity \cite{Ps,Recollisions}. Note that the magnetic field-induced ponderomotive drift motion into the forward direction is identical for the electron and positron due to the equal magnitude of their charge-to-mass ratios. In this respect the Ps system is distinguished from ordinary atoms where the heavy nucleus (or ionic core) stays behind, so that laser-driven collisions are suppressed at high field intensity \cite{report}. 
In the case of strongly laser-driven Ps, the collisional impact parameter can be as small as the initial spatial separation of the particles which is determined by the Ps Bohr radius $a_0\approx 1\,$\AA. This way, high particle current densities and collision luminosities are attainable \cite{collider,muonPLB}. The combination of Ps atoms with intense laser fields might therefore be considered as an ``$e^+e^-$ micro-collider''.
The threshold laser intensity to render muon production with significant probability from Ps or a low-energy $e^+e^-$ plasma is determined by the relation $e|\mbox{\boldmath$A$}|\ge Mc^2$, with the laser vector potential $\mbox{\boldmath$A$}$, because the collision energy comes from the transversal motion. For a near-infrared laser wavelength of $\lambda =1$ $\mu$m, the threshold intensity amounts to $I=5.5\times 10^{22}$\,W/cm$^2$, which is almost reached by the most powerful present laser systems \cite{Bahk,Uggerhoj}.

The process $e^+e^-\to\mu^+\mu^-$ in a laser field has also been discussed within a two-step mechanism recently: In Ref.~\cite{Kuchiev}, an $e^+e^-$ pair is assumed to be created in the collision of a laser pulse with a relativistic ion beam, and the muons are subsequently produced in a laser-driven $e^+e^-$ collision. The laser field strength in the ionic rest frame needs to approach the Schwinger value $E_{\rm cr}^{(e)} = m^2c^3/e\hbar=1.3\times 10^{16}$\,V/cm ($I\sim 10^{29}$\,W/cm$^2$). In this way, the well-established analogy between atomic ionization and $e^+e^-$ pair creation in strong laser fields has been extended to include also the recollision step.
In a broader sense, the investigation of lepton-lepton interactions in laser fields has a long history, comprising laser-assisted $e^-e^-$ (M{\o}ller) \cite{Moller}, $e^+e^-$ (Bhabha) \cite{Bhabha}, and $e^-\mu^-$ scattering \cite{electron-muon}. In contrast to the process under consideration here, these lepton-lepton scattering processes can also take place without a background laser field, and the presence of the latter merely modifies the field-free cross-section.

In the present paper, we give a comprehensive account of the reaction ${\rm Ps} \to \mu^+\mu^-$ in a laser wave of linear polarization, this way extending our recent study in Ref.\,\cite{muonPLB}. The Ps atom is supposed to be initially at rest and in its ground state. The $\mu^+\mu^-$ pair is created in the laser-driven $e^+e^-$ collisions as described above (see also Fig.\,1). On the one hand, we perform a quantum electrodynamical calculation to obtain the total production rate and muon spectra; both analytical and numerical results are presented. The occurence of sums of generalized Bessel functions of very high order $\gtrsim 10^{10}$ constitutes a major difficulty of the case of a linearly polarized driving field. On the other hand, 
particular emphasis is laid upon the derivation of a semi-classical model for the muon production process, representing one of the main extensions of \cite{muonPLB}. The model allows to interpret the typical muon momenta and the total process probability in a simple and intuitive manner. It is proven by a saddle-point consideration and resembles the famous three-step model of laser-driven recollisions \cite{Recollisions}, but also exhibits characteristic differences. The saddle-point treatment moreover provides us with a suitable asymptotic expansion to handle the high-order generalized Bessel functions.
Furthermore, a comparison is drawn with muon pair creation from Ps in a laser field of circular polarization, which was treated in Ref.~\cite{muonPRD}. In this situation it was found that the muon production is essentially suppressed, because the $e^+e^-$ rescattering occurs at large impact parameters. The latter are determined mainly by the classical electron and positron trajectories in the circularly polarized field and have the macroscopic dimension $\varrho\sim\lambda \xi\sim 10^{-2}$\,cm, where $\xi\sim 200$ is the laser intensity parameter [see Eq.\,(\ref{xi}) below]. Instead, in a linearly polarized field the collisional impact parameter can be microscopically small, $\varrho\sim a_0\approx 10^{-8}$\,cm, which enhances the probability for Ps decay into a muon pair by several orders of magnitude. Finally, the experimental feasibilty of laser-induced muon production is discussed in detail. It is shown that an observation of the process will be rendered possible in a crossed-beam setup employing high-power laser sources of the next generation combined with high-density Ps samples.

\begin{figure}[b]
\begin{center}
\resizebox{4.5cm}{!}{\includegraphics{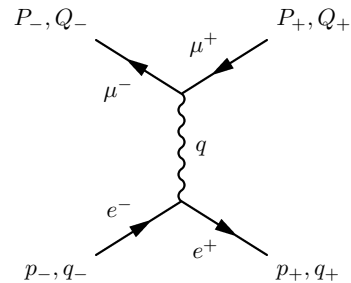}}
\caption{\label{diagramm} Feynman graph for muon pair creation from electron-positron annihilation in a background laser field. The arrows are labeled 
by the particle's free momenta ($p_\pm, P_\pm$) outside and the effective momenta 
($q_\pm, Q_\pm$) inside the laser field. The virtual photon has four-momentum $q$. The electron and positron are assumed to form a Ps atom initially.} 
\end{center} 
\end{figure}

The paper is organized as follows. In Sec.\,II we derive general expressions for the probability of laser-driven Ps decay into muons in two ways. First, we proceed in Sec.\,II.A along the lines of our earlier paper \cite{muonPRD} and evaluate the process amplitude by virtue of a Fourier expansion, where the Fourier coefficients are given by generalized Bessel functions. In Sec.\,II.B we employ an alternative approach based on the saddle-point integration method, which will prove useful in various respects: We show that important features of the considered muon production process can be understood in classical terms (see also Sec.\,III.B) and derive an asymptotic expansion of the generalized Bessel function [see Eq.\,(\ref{genbesasy})], which later serves us in our numerical computations. These special functions frequently occur in calculations of intense laser-matter interactions \cite{SFA3,Leubner,genbes2}. Based on the general expressions at hand, the relevant physical features of the process are elaborated in Sec.\,III. The kinematics of the produced muons is analyzed in Sec.\,III.A and explained in Sec.\,III.B within an intuitive semi-classical model. In Sec.\,III.C we derive a closed analytical approximation to the total creation rate [see Eq.\,(\ref{RPs2})], which exhibits the main physical features of the process and can be interpreted in terms of the cross section for the field-free reaction $e^+e^-\to\mu^+\mu^-$ and the quantum mechanical
dispersion of the initially bound $e^+e^-$ wave packet in the laser field. Afterwards, we briefly consider in Sec.\,III.D the related process of muon pair production by a superstrong laser wave of linear polarization which interacts with a nonrelativistic $e^+e^-$ plasma. The influence of the laser polarization is revealed in Sec.\,III.E.
In the following Sec.\,IV we present results of numerical calculations, compare them with previous ones for circular laser polarization, and discuss their relevance for an experimental investigation of laser-driven muon creation. Here, also more complex driving field configurations and their potential advantages are described. We summarize our main results and conclusions in Sec.\,V.

Relativistic units with $\hbar = c = 1$ are used throughout, except where otherwise stated. The finestructure constant is $\alpha=e^2\approx 1/137$. We employ the metric tensor $g^{\mu\nu}={\rm diag}(+---)$, so that the scalar product of two four-vectors $p^\mu=(p^0,\mbox{\boldmath$p$})$ and $q^\mu=(q^0,\mbox{\boldmath$q$})$ reads $(pq) = p^0q^0-\mbox{\boldmath$pq$}$. Furthermore, Feynman slash notation $\slash\!\!\!p = (\gamma p)$ is used, with the Dirac matrices $\gamma^\mu$.

\section{Theoretical framework}

\subsection{Transition amplitude and reaction rate}
The rate for positronium decay into muons in a strong laser field can be calculated within the framework of laser-dressed quantum electrodynamics (QED). The driving laser field is assumed to be a monochromatic plane wave of linear polarization, described by the classical four-potential
\begin{eqnarray}
\label{A}
A^\mu(x) = \epsilon^\mu a\cos(kx),
\end{eqnarray}
with the polarization four-vector $\epsilon^\mu=(0,1,0,0)$, the wave four-vector
$k^\mu=\omega(1,0,0,1)$, and the amplitude $a$. In the following consideration, the laser parameters have typical values of $\omega \sim 1\, {\rm eV}$ and $a\gtrsim 100$\,MV, corresponding to a high intensity $I\gtrsim 10^{22}$ W/cm$^2$ in the near-infrared frequency domain. When an electron and a positron are exposed to such an intense laser wave, the interparticle Coulomb interaction is much weaker than the coupling to the external field and may be neglected. Within the spirit of the strong-field approximation \cite{SFA1,SFA3}, the amplitude of the process ${\rm Ps}\to\mu^+\mu^-$ can therefore be expressed as a superposition integral \cite{muonPRD}
\begin{eqnarray}
\label{SPs}
\mathcal{S}_{{\rm Ps}} = {1\over\sqrt{\mathcal{V}}}\int{d^3p\over(2\pi)^3}\,
           \tilde\Phi(\mbox{\boldmath$p$})\, 
           \mathcal{S}_{e^+e^-}(\mbox{\boldmath$p$}),
\end{eqnarray}
with a normalization volume $\mathcal{V}$ and the Fourier transform  $\tilde\Phi(\mbox{\boldmath$p$})=
8\sqrt{\pi}a_0^{3/2}/(1+a_0^2\mbox{\boldmath$p$}^2)^2$ of the Ps ground state of Bohr radius $a_0$. The latter function is also known as the atomic Compton profile. The Ps atom is initially assumed at rest, and $\mbox{\boldmath$p$}$ denotes the relative momentum of the electron-positron two-body system. Equation\,(\ref{SPs}) describes the coherent average over the amplitude
\begin{eqnarray}
\label{See}
\mathcal{S}_{e^+e^-}(\mbox{\boldmath$p$}) &=& -{\rm i}\alpha\int d^4x \int d^4y
\overline\Psi_{p_+,s_+}(x)\gamma^\mu\Psi_{p_-,s_-}(x) \nonumber\\
& & \times D_{\mu\nu}(x-y)
\overline\Psi_{P_-,S_-}(y)\gamma^\nu\Psi_{P_+,S_+}(y)
\end{eqnarray}
for the process $e^+e^-\to\mu^+\mu^-$ in a laser wave (cf. Fig.\,1). The initial electron and positron momenta are $\mbox{\boldmath$p$}_\pm=\pm\mbox{\boldmath$p$}$. Their wave functions in Eq.\,(\ref{See}) are given by the Volkov states \cite{Vol,LL}
\begin{eqnarray}
\label{Vol}
\Psi_{p_\pm,s_\pm}(x)= \sqrt{m\over q_\pm^0}
\left(1\pm {e\slash\!\!\!k\slash\!\!\!\!A\over 2(kp_\pm)}\right)
u_{p_\pm,s_\pm}\,{\rm e}^{{\rm i}f^{(\pm)}},
\end{eqnarray}
with the phase
\begin{eqnarray*}
f^{\scriptscriptstyle{(\pm)}} = 
\pm (q_\pm x) + {ea(\epsilon p_\pm)\over (kp_\pm)}\sin(kx) \pm {e^2a^2\over 8(kp_\pm)}\sin[2(kx)].
\end{eqnarray*}
In Eq.\,(\ref{Vol}), $p_\pm^\mu=(p_0,\pm\mbox{\boldmath$p$})$ are the initial free four-momenta of the electron and positron (outside the laser field), $s_\pm$ denote the particle spin states, the $u_{p_\pm,s_\pm}$ are free Dirac spinors \cite{BD}, and 
\begin{eqnarray}
\label{q}
q_\pm^\mu = p_\pm^\mu +{e^2a^2\over 4(kp_\pm)}k^\mu
\end{eqnarray}
are the effective four-momenta of the particles inside the laser field. In accordance with Eq.\,(\ref{q}), the momentum vectors satisfy the relations
$\mbox{\boldmath$q$}_\pm^\perp = \mbox{\boldmath$p$}_\pm^\perp$ and
$\mbox{\boldmath$q$}_+^\perp + \mbox{\boldmath$q$}_-^\perp = 0$,
where the label $\perp$ denotes the component perpendicular 
to the laser propagation direction. The corresponding effective mass reads
$m_*^2 = q_\pm^2 = m^2(1 + \xi^2)$,
with the dimensionless laser intensity parameter
\begin{eqnarray}
\label{xi}
\xi = {ea\over m\sqrt{2}}.
\end{eqnarray} 
The Volkov states in Eq.\,(\ref{Vol}) are normalized to a number density of one particle per unit volume, and to a $\delta$-function in $q_\pm$ space \cite{LL,SZ}. Analogous expressions hold for the Volkov states $\Psi_{P_\pm,S_\pm}$, the free momenta $P_\pm^\mu$, the spin states $S_\pm$, the effective momenta $Q_\pm^\mu$, and the effective mass $M_\ast=(M^2+m^2\xi^2)^{1/2}$ of the muons. Moreover, we apply in Eq.\,(\ref{See}) the free photon propagator
\begin{eqnarray}
\label{D}
D_{\mu\nu}(x-y) = \int {d^4q\over(2\pi)^4}
                  {{\rm e}^{{\rm i}q\cdot(x-y)}\over q^2}g^{\mu\nu}.
\end{eqnarray}

We comment on various approximations that are implicitly contained in the above equations. (1) The initial spin coupling in the Ps ground-state is neglected in Eq.\,(\ref{SPs}). It is of minor importance in the presence of the strong laser wave; the corresponding hyperfine splitting between the ortho- and para-states amounts to less than 1\,meV which is negligibly small on the energy scale set by the external field. Moreover, later on we will average over the $e^\pm$ spin states [see Eq.\,(\ref{RPs})] so that our approach applies to a statistical ensemble of Ps atoms. A spin state-resolved calculation could be performed by inserting appropriate projection operators \cite{spinproj}. (2) The amplitude in Eq.\,(\ref{See}) fully accounts for the interaction of the leptons with the laser field, while their interaction with the QED vacuum is taken into account to lowest order, in accordance with the Feynman graph in Fig.\,1. The interplay between the Coulomb field and the electromagnetic wave during the very beginning of the laser field's rising edge might lead to small distortions in the momentum distribution $\tilde\Phi(\mbox{\boldmath$p$})$. (3) Our treatment moreover neglects the radiative energy loss of the laser-driven electron and positron due to Thomson scattering, which is rather small for the laser parameters under consideration \cite{muonPRD}. (4) Finally, we ignore in Eq.\,(\ref{D}) modifications of the photon propagator due to the presence of the external field. These modifications become important only at substantially higher laser intensities $I\gtrsim 10^{29}$ W/cm$^2$ \cite{prop,Erik}.

The space-time integrations in Eq.\,(\ref{See}) can be performed by the standard method of Fourier series expansion, using the generating function of the generalized Bessel functions \cite{SFA3}. The latter can be expressed via ordinary Bessel functions according to \cite{SFA3}
\begin{eqnarray}
\label{genbes}
J_n(u,v) = \sum_{\ell = -\infty}^\infty J_{n-2\ell}(u)J_\ell (v).
\end{eqnarray}
For the amplitude, we obtain this way
\begin{eqnarray}
\label{See1}
\mathcal{S}_{e^+e^-}(\mbox{\boldmath$p$}) &=& -{\rm i} (2\pi)^4 \alpha
{m\over\sqrt{q_+^0 q_-^0}}{M\over\sqrt{Q_+^0 Q_-^0}} \nonumber\\
&\times&\!\! \int {d^4q\over q^2} \sum_{n,N} \mathcal{M}^\mu(p_+,p_-|n) \mathcal{M}_\mu(P_+,P_-|N) \nonumber\\
&\times&\!\! \delta(q_+ + q_- -q -nk)\, \delta(Q_+ + Q_- -q -Nk),\nonumber\\
& &
\end{eqnarray}
with the spinor-matrix products
\begin{widetext}
\begin{eqnarray}
\label{M}
\mathcal{M}^\mu(p_+,p_-|n) =
\bar u_{p_+,s_+}\bigg\{\gamma^\mu b_n 
+\left({ea\slash\!\!\!\epsilon\slash\!\!\!k\gamma^\mu\over 2(kp_+)}
      -{ea\gamma^\mu\slash\!\!\!k\slash\!\!\!\epsilon\over 2(kp_-)} \right)c_n
- {e^2a^2 k^\mu\slash\!\!\!k\over 2(kp_+)(kp_-)}d_n
\bigg\} u_{p_-,s_-}\,, \nonumber\\
\mathcal{M}_\mu(P_+,P_-|N) =
\bar u_{P_-,S_-}\bigg\{\gamma_\mu B_N 
-\left({ea\slash\!\!\!\epsilon\slash\!\!\!k\gamma_\mu\over 2(kP_-)}
      -{ea\gamma_\mu\slash\!\!\!k\slash\!\!\!\epsilon\over 2(kP_+)} \right)C_n
- {e^2a^2 k_\mu\slash\!\!\!k\over 2(kP_+)(kP_-)}D_n
\bigg\} u_{P_+,S_+}.
\end{eqnarray}
\end{widetext}
The coefficients of the electronic spinor-matrix products read
\begin{eqnarray*}
\label{coeff1}
b_n &=& J_n(u,v)\,, \nonumber \\
c_n &=& {1\over 2}\,\left[ J_{n-1}(u,v) + J_{n+1}(u,v) \right]\,, \nonumber\\
d_n &=& {1\over 4}\,\left[ J_{n-2}(u,v) + 2J_n(u,v) + J_{n+2}(u,v) \right]
\end{eqnarray*}
with the arguments
\begin{eqnarray}
\label{arguv}
u &=& {ea(\epsilon p_-)\over (kp_-)}-{ea(\epsilon p_+)\over (kp_+)}\,, \nonumber\\
v &=& -{e^2a^2\over 8}\left[{1\over (kp_-)} + {1\over (kp_+)}\right].
\end{eqnarray}
Note that $v\approx v_0\equiv -m\xi^2/(2\omega)$ is practically constant for the 
relevant values of $|\mbox{\boldmath$p$}|\sim m\alpha$.
The coefficients $B_N$, $C_N$ and $D_N$ of the muonic spinor-matrix products
have the same structure, with the arguments given by
\begin{eqnarray}
\label{argUV}
U &=& {ea(\epsilon P_-)\over (kP_-)}-{ea(\epsilon P_+)\over (kP_+)}\,, \nonumber\\
V &=& -{e^2a^2\over 8}\left[{1\over (kP_-)} + {1\over (kP_+)}\right].
\end{eqnarray}
According to the energy-momentum conserving $\delta$-function at the first vertex of the Feynman graph, the integer number $n$ in Eq.\,(\ref{See1}) 
counts the laser photons that are emitted (if $n>0$) or absorbed (if $n<0$) by the electron and positron. Similarly, $N$ is the number of photons exchanged with the laser field by the muons. Denoting the total number of absorbed laser photons by $r:=N-n$ and integrating over the virtual photon momentum yields
\begin{eqnarray}
\label{See2}
\mathcal{S}_{e^+e^-}(\mbox{\boldmath$p$}) &=& -{\rm i} (2\pi)^4 \alpha
{m\over\sqrt{q_+^0 q_-^0}}{M\over\sqrt{Q_+^0 Q_-^0}} \nonumber\\
& &\times\sum_{n,r} \mathcal{M}^\mu(p_+,p_-|n) \mathcal{M}_\mu(P_+,P_-|n+r)\nonumber\\
& &\times
{\delta(q_+ + q_- - Q_+ - Q_- + rk)\over (q_+ + q_- - nk)^2}.
\end{eqnarray}
By Eq.\,(\ref{genbes}) and the properties of the ordinary Bessel
functions \cite{AS}, the main contribution to the amplitude comes from 
photon numbers $n\approx 2v\approx 2v_0 = -m\xi^2/\omega$.
Consequently, the square of the virtual photon momentum
\begin{eqnarray}
\label{qsq}
 (q_+ + q_- - nk)^2 &=& 2[m_\ast^2+(q_+q_-)-n(kp_+)-n(kp_-)] \nonumber\\
  &\approx& 4(m_\ast^2 - n\omega m) \approx 8m^2\xi^2
\end{eqnarray}
is practically constant and may be pulled out of the summation over $n$ in Eq.\,(\ref{See2}):
\begin{eqnarray}
\label{See3}
\mathcal{S}_{e^+e^-}(\mbox{\boldmath$p$}) &\approx& -{\rm i} (2\pi)^4 \alpha
{m\over\sqrt{q_+^0 q_-^0}}{M\over\sqrt{Q_+^0 Q_-^0}} {1\over 8m^2\xi^2} \nonumber\\
& &\times \sum_{n,r} \mathcal{M}^\mu(p_+,p_-|n) \mathcal{M}_\mu(P_+,P_-|n+r)\nonumber\\
& &\times \delta(q_+ + q_- - Q_+ - Q_- + rk).
\end{eqnarray}
A particular implication of Eq.\,(\ref{qsq}) is that the process proceeds nonresonantly (i.e. $q^2\ne 0$) \cite{resonances}.
For later use we note further that the typical energy of the virtual photon is
\begin{eqnarray}
\label{q0}
q_0 = q_+^0 + q_-^0 - n\omega \approx 2m\xi^2.
\end{eqnarray}
The sum over $n$ in Eq.\,(\ref{See3}) can be performed analytically by 
virtue of an addition theorem for the generalized Bessel functions \cite{SFA3} 
\begin{eqnarray}
\label{addthm}
\sum_{n = -\infty}^\infty J_n(u,v)J_{n+r}(U,V) = J_r(U-u,V-v).
\end{eqnarray}
In this manner we find
\begin{eqnarray}
\label{uvw}
& & \sum_n \mathcal{M}^\mu(p_+,p_-|n)\mathcal{M}_\mu(P_+,P_-|n+r) 
= J_r u^\mu U_\mu \nonumber\\
& & + K_r (u^\mu V_\mu + v^\mu U_\mu)
+ L_r (u^\mu W_\mu + v^\mu V_\mu + w^\mu U_\mu) \nonumber\\ 
& & +\ M_r (v^\mu W_\mu + w^\mu V_\mu) + N_r w^\mu W_\mu.
\end{eqnarray}
Here we have introduced the notation
\begin{eqnarray*}
\label{spinor-matrix-products}
u^\mu &=& \bar u_{p_+,s_+}\gamma^\mu u_{p_-,s_-} \nonumber\\
v^\mu &=& \bar u_{p_+,s_+}
\left({ea\slash\!\!\!\epsilon\slash\!\!\!k\gamma^\mu\over 2(kp_+)}
      -{ea\gamma^\mu\slash\!\!\!k\slash\!\!\!\epsilon\over 2(kp_-)} \right)
u_{p_-,s_-} \nonumber \\
w^\mu &=& \bar u_{p_+,s_+}
\left(- {e^2a^2 k^\mu\slash\!\!\!k\over 2(kp_+)(kp_-)}\right)
u_{p_-,s_-}
\end{eqnarray*}
and similarly $U_\mu, V_\mu$, and $W_\mu$ for the muons. 
The coefficients in Eq.\,(\ref{uvw}) read
\begin{eqnarray*}
\label{coeff2}
J_r &=& J_r(U-u,V-v), \nonumber\\
K_r &=& {1\over 2}\left( J_{r-1} + J_{r+1} \right),\nonumber \\
L_r &=& {1\over 4}\left( J_{r-2} + 2J_r + J_{r+1} \right),\nonumber \\
M_r &=& {1\over 8}\left( J_{r-3} + 3J_{r-1} + 3J_{r+1} + J_{r+3} \right),\nonumber \\
N_r &=& {1\over16}\left( J_{r-4} + 4J_{r-2} + 6J_r + 4J_{r+2} + J_{r+4}\right).
\end{eqnarray*}

After these transformations of the amplitude for the process $e^+e^-\to\mu^+\mu^-$ in Eq.\,(\ref{See}) we come back to the amplitude for the decay ${\rm Ps}\to\mu^+\mu^-$ in Eq.\,(\ref{SPs}). We need to multiply Eq.\,(\ref{See3}) by the Compton profile $\tilde\Phi(\mbox{\boldmath$p$})$ of the Ps ground state and integrate over the relative momentum $\mbox{\boldmath$p$}$. This complicated integral can be performed in an approximate way as follows:

Within the momentum range given by $\tilde\Phi(\mbox{\boldmath$p$})$, the electronic spinor-matrix products in Eq.\,(\ref{spinor-matrix-products}) may be considered as constant and pulled out of the integral. The same holds for the kinematic factors $q_\pm^0\approx m\xi^2/2$ and the energy-momentum conserving $\delta$-function. (The latter is, in fact, a subtle issue and dealt with in Appendix A). Hence, we are left with integrals of the form
\begin{eqnarray}
\label{int}
\bar{J_r} = \int {d^3p\over(2\pi)^3} \tilde\Phi(\mbox{\boldmath$p$})J_r(U-u,V-v).
\end{eqnarray}
Despite the small $\mbox{\boldmath$p$}$ values contained in the Compton 
profile, the high-order Bessel function [cf. Eq.\,(\ref{barr}) below] in the 
integrand is heavily oscillating, which leads to a small value of $\bar{J_r}$. 
The oscillatory damping of the amplitude is due to a destructive interference 
of the various partial waves within the Ps wave packet, which can equivalently 
be described as wave packet spreading [see Eq.\,(\ref{RPs3})]. The evaluation of the integral (\ref{int}) is performed in Appendix B. With $\Delta V\equiv V-v_0$, the result reads 
\begin{eqnarray}
\label{barJr}
\bar{J_r} \approx \sqrt{2^9\over\pi^3a_0^3}\left({\omega\over m\alpha^2\xi}\right)^{2/3}
{(\alpha\xi\Delta V)^{1/3}\over\alpha^2|v_0|} J_r(U,\Delta V).
 \end{eqnarray}

Now we can write the square of the Ps decay amplitude as
\begin{eqnarray}
\label{Ssq}
|\mathcal{S}_{\rm Ps}|^2 &=& (2\pi)^4 \alpha^2
{m^2\over q_+^0 q_-^0}{M^2\over Q_+^0 Q_-^0} {1\over 64 m^4\xi^4} \nonumber\\
& & \times\sum_r \left|\overline{\sum_n} \mathcal{M}^\mu(p_+,p_-|n) \mathcal{M}_\mu(P_+,P_-|n+r) \right|^2\nonumber\\
& &\times\ \delta(q_+ + q_- - Q_+ - Q_- + rk) \mathcal{V}T\,,
\end{eqnarray}
where the symbol $\overline{\sum}_n$ indicates both the sum over $n$ and the average
over the Ps ground state, as described above. $p_\pm$ and $q_\pm$ in Eq.\,(\ref{Ssq}) are to be understood as some typical values of the electron and positron momenta. The 
factors of volume $\mathcal{V}$ and time $T$ come, as usual, from the square of the $\delta$-function. Note that the energy-momentum conserving $\delta$-function implies $\mbox{\boldmath$Q$}_+^\perp + \mbox{\boldmath$Q$}_-^\perp =0$.
From Eq.\,(\ref{Ssq}) we obtain the total rate for laser-driven Ps decay into muons by averaging over the initial spin states, summing over the final spin states, and integrating over the final momenta:
\begin{eqnarray}
\label{RPs}
R_{\rm Ps} = {1\over T}
\int {d^3 Q_+\over (2\pi)^3} \int {d^3 Q_-\over (2\pi)^3}\, 
{1\over 4} \sum_{s_\pm,S_\pm}|\mathcal{S}_{\rm Ps}|^2.
\end{eqnarray}
Based on Eq.\,(\ref{RPs}), we have performed numerical calculations; the results will be shown in Sec.\,IV. It is also possible to evaluate the integrals in Eq.\,(\ref{RPs}) in an approximate way by analytical means. This derivation is performed in Sec.\,III.C.

\subsection{Saddle-point approach}
Due to the periodicity of the laser field, the exponentials in the amplitude 
(\ref{See}) are highly oscillating functions. As an alternative to the series 
expansion into generalized Bessel functions as performed in Eq.\,(\ref{See1}), one can 
evaluate the amplitude by applying the saddle-point integration method. Our aim here is twofold: on the one hand, to derive a simplified expression for the Ps decay amplitude (\ref{Ssq}), and on the other hand, to interpret the saddle-point integration results via a simple-man model which will lead to an intuitive interpretation of the process probability (see Sec.\,III.B).

It is convenient to introduce light-like 
coordinates $x_\pm\equiv x^0\pm x^3$ and $y_\pm\equiv y^0\pm y^3$, so that the 
integrals over $dx_+$, $dx^\perp$, $dy_+$ and $dy^\perp$ all result in $\delta$
functions. Among them are $\delta(q_+^\perp+q_-^\perp-q^\perp)=\delta(q^\perp)$
and $\delta(q_+^0-q_+^3+q_-^0-q_-^3-q^0+q^3)=\delta(2p_0-q^0+q^3)$, which we 
exploit to integrate by $dq^\perp$ and $dq^3$. This yields 
\begin{eqnarray}
\label{SeeSP}
\mathcal{S}_{e^+e^-} &=& -{\rm i}\alpha \pi^2 \int dx_- \int dy_-
\int {dq_0\over p_0(q_0-p_0)} \nonumber\\
& & \times {\rm e}^{-{\rm i}(q_+^0+q_-^0-q_0)x_-+{\rm i}u\sin(\omega x_-)+{\rm i}v\sin(2\omega x_-)} \nonumber\\
& & \times {\rm e}^{ {\rm i}(Q_+^0+Q_-^0-q_0)y_--{\rm i}U\sin(\omega y_-)-{\rm i}V\sin(2\omega y_-)} \nonumber\\
& & \times {\mathcal M}^\mu(p_+,p_-;x_-){\mathcal M}_\mu(P_+,P_-;y_-)\nonumber\\
& & \times \delta(Q_+^\perp+Q_-^\perp)\delta(Q_+^0-Q_+^3+Q_-^0-Q_-^3-2p_0) \nonumber\\
& &
\end{eqnarray}
with 
\begin{eqnarray}
\label{Mppx}
& & \!\!\!\!\!\!  {\mathcal M}^\mu(p_+,p_-;x_-) \equiv {m\over\sqrt{q_+^0 q_-^0}} \nonumber\\
& & \!\!\!\!\!\!   \times\, {\bar u}_{p_+,s_+}\left(1+{e\slash\!\!\!\!A\slash\!\!\!k\over 2(kp_+)}\right)
\gamma^\mu
\left(1-{e\slash\!\!\!k\slash\!\!\!\!A\over 2(kp_-)}\right)u_{p_-,s_-}
\end{eqnarray}
and accordingly ${\mathcal M}_\mu(P_+,P_-;y_-)$ for the muons.
The integrals over $dx_-$ and $dy_-$ can separately be done by the saddle-point 
method. We explicitly demonstrate the $y_-$ integration by considering first the term in ${\mathcal M}_\mu(P_+,P_-;y_-)$ which is proportional to ${\bar u}_{P_-,S_-} \gamma^\mu u_{P_+,S_+}$ and independent of $y_-$. The corresponding integral is denoted as
\begin{eqnarray}
\label{I}
\mathcal{I} \equiv \int_{-\infty}^{+\infty} dy_-\,{\rm e}^{ {\rm i}f(y_-)},
\end{eqnarray}
with the phase given by
\begin{eqnarray}
\label{f}
f(y_-)\equiv \Delta Q y_- - U\sin(\omega y_-) - V\sin(2\omega y_-),
\end{eqnarray}
where $\Delta Q\equiv Q_+^0 + Q_-^0 - q_0$. The saddle point equation
\begin{eqnarray}
\label{SPE}
{df\over dy_-} =
\Delta Q - U\omega\cos(\omega y_-) - 2V\omega\cos(2\omega y_-) = 0
\end{eqnarray}
has the solution
\begin{eqnarray}
\label{cosy}
\cos(\omega y_-) = -{U\over 8V} \pm 
\sqrt{{U^2\over 64V^2}+{\Delta Q\over 4V\omega}+{1\over 2}}\,.
\end{eqnarray}
The resulting saddle points $y_-$ indicate the regions giving the main contribution to the process. Physically speaking, they determine the laser phases when the muons are created predominatly. The corresponding saddle points of the $x_-$ integral represent, accordingly, the laser phases when the $e^+e^-$ annihilation occurs. This fact will be utilized in Sec.\,III.B below.

Due to the periodicity of the phase (\ref{f}), we can rewrite Eq.\,(\ref{I}) as
\begin{eqnarray}
\label{Ialtern}
\mathcal{I} &=& \sum_{N=-\infty}^{+\infty}\int_{2\pi N}^{2\pi(N+1)} {d\theta\over\omega}\, 
{\rm e}^{{\rm i}(U\sin\theta+V\sin2\theta-\nu\theta)} \nonumber\\
&=& {1\over\omega}\sum_{N=-\infty}^{+\infty}{\rm e}^{-2\pi {\rm i} N\frac{\Delta Q}{\omega}}
\int_0^{2\pi} d\theta' {\rm e}^{{\rm i}(U\sin\theta'+V\sin2\theta'-\nu\theta')} \nonumber\\
&=& 2\pi\sum_{N=-\infty}^{+\infty}\delta(\Delta Q-N\omega) J_N(U,V)\,,
\end{eqnarray}
with $\nu\equiv\Delta Q/\omega$, $\theta\equiv -\omega x_-$ and 
$\theta'\equiv \theta-2\pi N$. In the last step, the integral representation $J_N(U,V) = (1/2\pi) \int_0^{2\pi} \exp[{\rm i}(U\sin\theta+V\sin2\theta-N\theta)] d\theta$
from Ref.\,\cite{SFA3} has been applied. If the $\theta'$ integral is evaluated instead with the help of the saddle points in Eq.\,(\ref{cosy}), one obtains the asymptotic expansion
\begin{widetext}
\begin{eqnarray}
\label{genbesasy}
J_N(U,V) = \sqrt{1\over 4\pi\rho}\left(
{\cos\left(N\eta_+ - U\sin\eta_+ - V\sin 2\eta_+ +\sigma^{(+)} {\pi\over 4}\right)
\over\sqrt{|V\sin\eta_+|}} +
{\cos\left(N\eta_- - U\sin\eta_- - V\sin 2\eta_- -\sigma^{(-)} {\pi\over 4}\right)
\over\sqrt{|V\sin\eta_-|}} \right),
\end{eqnarray}
\end{widetext}
provided that the saddle points are real and isolated from each other. Here we have introduced the notation
\begin{eqnarray*}
\rho &=& \sqrt{{U^2\over 64V^2}+{N\over 4V}+{1\over 2}}\,,\\
\eta_\pm &=& \arccos\left(-{U\over 8V} \pm \rho\right)\,,\\
\sigma^{(\pm)} &=& {\rm sgn}(V\sin\eta_\pm).
\end{eqnarray*}
The asymptotic representation (\ref{genbesasy}) has been used in our numerical calculations (see Sec.\,IV). It considerably simplifies the evaluation of high-order generalized Bessel functions in comparison with the series expansion (\ref{genbes}). For the reader's convenience, the details of the derivation are given in Appendix C. The formula (\ref{genbesasy}) is also contained as Eq.\,(3.8) in the comprehensive study of asymptotic expansions of generalized Bessel functions in Ref.\,\cite{Leubner}. Its application presumes real parameters $U$, $V$, and $N$ of large magnitude. The range of applicability is specified more precisely in the appendix.

We briefly check that the present approach leads to the same result as in Sec.\,II.A, where first the integrations over $d^4x$ and $d^4y$ have been performed and afterwards the $d^4q$ integral. By using Eq.\,(\ref{Ialtern}), the expression 
\begin{eqnarray}
\label{SeeSP2}
\mathcal{S}_{e^+e^-}\!\! &\sim&\!\! -{\rm i}\alpha 4\pi^4 
\int {dq_0\over p_0(q_0-p_0)+{\rm i}\epsilon}
\sum_n \delta(\Delta q-n\omega) \nonumber\\
& & \times J_n(u,v) \sum_N \delta(\Delta Q-N\omega) J_N(U,V)
\nonumber\\
& & \times
{m\over\sqrt{q_+^0 q_-^0}} {M\over\sqrt{Q_+^0 Q_-^0}}\, u^\mu U_\mu\,
\delta(Q_+^\perp+Q_-^\perp)\nonumber\\
& & \times\delta(Q_+^0-Q_+^3+Q_-^0-Q_-^3-2p_0),
\end{eqnarray}
with $\Delta q\equiv q_+^0 + q_-^0 - q_0$, is found for the leading term in Eq.\,(\ref{SeeSP}) which contains the integral (\ref{I}). The integration over $dq_0$ in Eq.\,(\ref{SeeSP2}) can be performed with the help of the first $\delta$ function 
yielding
\begin{eqnarray}
\label{SeeSP3}
\mathcal{S}_{e^+e^-}\!\! &\sim&\!\! -{\rm i}\alpha (2\pi)^4 
\sum_{n,N} {J_n(u,v) J_N(U,V)\over (q_+ + q_- - nk)^2} 
{m\over\sqrt{q_+^0 q_-^0}} {M\over\sqrt{Q_+^0 Q_-^0}}\nonumber\\
& & \times\, u^\mu U_\mu \,\delta^{(4)}\left(Q_+ + Q_- - q_+ - q_- -(N-n)k\right).\nonumber\\
\,
\end{eqnarray}
Here we used the relation
\begin{eqnarray*}
p_0(q_0-p_0) &\approx& m(q_+^0 + q_-^0 - n\omega -m) \\
&\approx&  {1\over 4}(q_+ + q_- - nk)^2
\end{eqnarray*}
which follows from Eq.\,(\ref{qsq}), and combined the remaining $\delta$ functions 
in Eq.\,(\ref{SeeSP2}) into a four-dimensional one by virtue of the identity
\begin{eqnarray*}
Q_+^0 + Q_-^0 &=& q_+^0 + q_-^0 + (N-n)\omega \nonumber\\
&=& 2m + q_{+,z} + q_{-,z} + (N-n)\omega.
\end{eqnarray*}
We note that the structure of Eq.\,(\ref{SeeSP3}) coincides with that of
Eq.\,(\ref{See2}).

So far, we have only dealt with the space-time independent parts of 
${\mathcal M}^\mu(p_+,p_-;x_-)$ and ${\mathcal M}_\mu(P_+,P_-;y_-)$ 
in Eq.\,(\ref{SeeSP}). There are also terms containing $\cos(\omega y_-)$ 
and $\cos^2(\omega y_-)$ in ${\mathcal M}_\mu(P_+,P_-;y_-)$ which can be handled
in a similar manner as before. For example, by writing 
$\cos(\omega y_-)={1\over 2}({\rm e}^{{\rm i}\omega y_-}+{\rm e}^{-{\rm i}\omega y_-})$,
the integral corresponding to Eq.\,(\ref{I}) becomes
\begin{eqnarray*}
\mathcal{I}_1 \equiv \int dy_-\,{1\over 2}
\left({\rm e}^{ {\rm i}f_+(y_-)} + {\rm e}^{ {\rm i}f_-(y_-)}\right),
\end{eqnarray*}
with the phases
\begin{eqnarray*}
f_\pm(y_-) &\equiv& (Q_+^0+Q_-^0-q^0\pm\omega)y_- - U\sin(\omega y_-) \nonumber\\
& & + 2V\sin(2\omega y_-).
\end{eqnarray*}
By recapitulating the previous calculation, we see that the new phases $f_\pm$ 
lead to the replacement of $\Delta Q$ by $\Delta Q\pm\omega$ or, 
correspondingly, of $N$ by $N\pm 1$. Thus, we obtain
\begin{eqnarray*}
\label{I1fin}
\mathcal{I}_1 = 2\pi\!\!\sum_{N=-\infty}^{+\infty}\!\!\delta(\Delta Q-N\omega) 
{1\over 2}\left[J_{N+1}(U,V)+J_{N-1}(U,V)\right]
\end{eqnarray*}
in analogy with the Bessel function treatment in Sec.\,II.A [cf. Eq.\,(\ref{coeff1})]. As expected, the saddle-point approach is equivalent to the treatment which employs a Fourier expansion into Bessel functions, when a suitable asymptotic representation of the latter is applied. For the analogous result in the case of a circularly polarized laser field, we refer to Ref.~\cite{LPHYS}.

\section{Analytical developments and results}

\subsection{Kinematical analysis}
By combining energy-momentum conservation as expressed by the $\delta$ function 
in Eq.\,(\ref{Ssq}) with Bessel function properties, we can gain rather 
comprehensive information on the kinematics of the produced muons (see \cite{muonPRD} 
for a similar analysis). The energy-momentum conservation condition
\begin{eqnarray*}
4m_\ast^2 + 4r\omega m \approx (q_+ + q_- +rk)^2 = (Q_+ + Q_-)^2\ge 4M_\ast^2
\end{eqnarray*}
implies that the minimal number of laser photons to be absorbed is
\begin{eqnarray}
\label{rmin}
r_{\rm min} = {M^2-m^2\over\omega m} \approx {M^2\over\omega m}.
\end{eqnarray} 
It is interesting to note that even this minimal photon number gives rise to ultrarelativistic muon momenta $Q_\pm^0\approx q_\pm^0 + r_{\rm min}\omega/2 \gg M_\ast$.
The $\delta$ function further yields
\begin{eqnarray*}
Q_+^0 - Q_{+,z} + Q_-^0 - Q_{-,z} &=& q_+^0 - q_{+,z} + q_-^0 - q_{-,z} \nonumber\\
&=& p_+^0 - p_{+,z} + p_-^0 - p_{-,z} \nonumber\\
&=& 2p_0 \approx 2m\,,
\end{eqnarray*}
which implies
\begin{eqnarray}
\label{kp}
(kP_+) \approx (kP_-) \approx (kp_+) \approx (kp_-) \approx \omega m
\end{eqnarray}
under the symmetry assumption $Q_+^0\approx Q_-^0$. The latter condition
will always be assumed in what follows, which allows us to simplify the notation by dropping the particle indices $\pm$ from time to time. From Eq.\,(\ref{kp}) we obtain 
in particular the relation
\begin{eqnarray}
\label{Vv}
 V \approx v
\end{eqnarray}
between the Bessel function arguments of Eqs.\,(\ref{arguv}) and (\ref{argUV}).
By expressing the virtual photon momentum in terms of the muon momenta
\begin{eqnarray*}
8m^2\xi^2 &\approx& (q_+ + q_- -nk)^2 = (Q_+ + Q_- -Nk)^2 \nonumber\\
&\approx& 4M_\ast^2 + 4Q_\perp^2 - 4N\omega m\,,
\end{eqnarray*}
we find that the typical number $r = N-n$ of absorbed laser photons is
\begin{eqnarray}
\label{barr}
\bar r \approx {M^2+Q_\perp^2\over\omega m}.
\end{eqnarray}
Due to Eq.\,(\ref{Vv}), the difference $\Delta V = V - v_0$ has a smaller
magnitude than $V$ or $v_0$ themselves and, consequently, than $r$. In order to 
obtain non-negligibly small values of the Bessel function $J_r(U,\Delta V)$ 
in Eq.\,(\ref{barJr}), it is therefore neccessary to have $U\approx r$. 
By Eq.\,(\ref{barr}) this implies
\begin{eqnarray}
\label{Px}
P_x = \sqrt{2}m\xi\left( 1 + \sqrt{1-\kappa^2}\right),
\end{eqnarray}
with $\kappa^2 \equiv (M^2+P_y^2)/(2m^2\xi^2)$. Requiring $\kappa^2\le 1$, we find 
\begin{eqnarray}
\label{Py}
|P_y|\le P_y^{\rm max}\equiv\sqrt{2}m\sqrt{\xi^2-\xi_{\rm min}^2}\ ,
\end{eqnarray} 
along with the minimal laser intensity parameter
\begin{eqnarray}
\label{ximin}
\xi_{\rm min}\equiv {M\over m\sqrt{2}}\,.
\end{eqnarray} 
We emphasize that $\xi_{\rm min}$ is minimal in the sense that it is the smallest $\xi$-value which allows for a non-negligibly small process pobability. In a background laser field the decay ${\rm Ps}\to\mu^+\mu^-$ is energetically possible at any laser intensity in principle, since the external field represents a practically infinite energy reservoir. However, only for laser intensities satisfying Eq.\,(\ref{ximin}) the probability of this reaction can be significant, as it guarantees a sufficiently efficient photon absorption. Note moreover that Eq.\,(\ref{ximin}) confirms the naive estimate of the threshold intensity given in the introduction. This is, in fact, remarkable since the muons have to be created in the field with their laser-dressed mass $M_\ast$, which is significantly larger than their bare mass $M$ (see also Sec.\,IV.B).

By combining Eqs.\,(\ref{barr})-(\ref{Py}) we find
\begin{eqnarray}
Q_z \approx {1\over 2}\left(q_{+,z} + q_{-,z} + \bar r \omega\right) 
\approx {M^2+3m^2\xi^2\over 2m}\ ,
\end{eqnarray}
which translates into 
\begin{eqnarray}
\label{Pz}
P_z \approx {M^2+2m^2\xi^2\over 2m}
\end{eqnarray}
by Eq.\,(\ref{q}). Close to threshold ($\xi\approx\xi_{\rm min}$), the produced 
muons have typical momenta of 
\begin{eqnarray}
\label{typicalP}
P_x \approx M\,,\ \ P_y \approx 0\,,\ \ P_z \approx {M^2\over m}. 
\end{eqnarray}
The total number of absorbed laser photons is $r\approx 2M^2/\omega m$, which is 
of the order of 10$^{10}$ at $\omega = 1$ eV and in agreement with the energy
conservation law outside the laser field: $r\omega \approx 2(P_0-m)$. Note that, 
even at threshold, the produced muons are highly relativistic such that their 
life time in the lab frame is considerably increased due to time dilation.
The typical muon momenta in Eq.\,(\ref{typicalP}) can be interpreted in terms
of a classical simple man's model as described in the following Sec.\,III.B.

For later use we note that the typical muon momenta given above imply, that the 
particles are emitted into a narrow cone along the laser propagation direction. 
Indeed one can show \cite{muonPRD}, that the opening angle of this cone is
\begin{eqnarray}
\label{thetamax}
\theta_{\rm max} \approx {2m\over M_\ast}\,.
\end{eqnarray}
The occurence of a maximum polar emission angle results from the fact that the electron and positron move at relativistic longitudinal velocity in the laboratory frame. In the c.m. frame, the muons are emitted isotropically. In contrast to the rotationally symmetric case of circular laser polarization \cite{muonPRD}, there is
also a maximum azimuthal angle. According to Eqs.\,(\ref{Px}) and (\ref{Py}),
we have $P_y^{\rm max}\lesssim P_x$ and thus $P_x\approx P_\perp$.
Consequently, 
\begin{eqnarray}
\label{phimax}
\phi_{\rm max} &=& \arcsin\left({P_y^{\rm max}\over P_\perp}\right) \nonumber\\
&\approx& \left\{\begin{array}{l}
   \displaystyle{ {P_y^{\rm max}\over P_\perp}
    \ \ \ ({\rm for}\ P_y^{\rm max}\ll P_\perp),}\\
   \\
   \displaystyle{\ \ {\pi\over 2}
   \ \ \ \ \ \ ({\rm for}\ P_y^{\rm max}\approx P_\perp)}
                 \end{array}\right. \nonumber\\
&\approx& {P_y^{\rm max}\over P_\perp} \approx \sqrt{1-{\xi_{\rm min}^2\over\xi^2}}\ .
\end{eqnarray}
The angle $\phi_{\rm max}$ determines the maximally allowed deviation from the laser 
polarization axis.

\subsection{Simple man's model of muon creation}
By using a classical simple-man's model, one can give an intuitive interpretation of the typical momenta in Eq.\,(\ref{typicalP}). This is similar to the famous three-step model of above-threshold ionization and high-harmonic generation in strong laser fields, which can also be derived within a saddle-point approach by equipping the saddle points with a classical meaning (see \cite{Becker} for a review).
In this simplified model the electron and positron are released from the bound Ps state by a strong laser field, excurse in the laser field as free particles acquiring energy from photon absorption, and produce a muon pair at the moment of the recollision. Let us consider the process at threshold ($\xi=\xi_{\rm min}$), where one may assume that the particles are created with zero momentum in the (primed) electron-positron c.m. frame, whenever the laser vector potential reaches its maximum amplitude. We note that the latter implies vanishing electric field strength of the laser. This agrees with the intuitive classical picture of laser-driven $e^+e^-$ (re)collisions, which always occur near zero electric field strength. It moreover indicates that the muon pairs are not produced directly by the laser field but rather by the energetic electron-positron collisions.

First, we demonstrate that the above assumptions indeed lead to Eq.\,(\ref{typicalP}). 
Let $\mbox{\boldmath$P$}'(\tau_0)=0$, where $\tau_0$ denotes the laser phase when the production occurs. The solution of the classical equations of motion with this initial condition reads
\begin{eqnarray}
\label{EoM}
P'_x(\tau) &=& e[A(\tau)-A(\tau_0)], \nonumber\\
P'_y(\tau) &=& 0, \nonumber\\ 
P'_z(\tau) &=& {e^2\over 2M}[A(\tau)-A(\tau_0)]^2,
\end{eqnarray} 
where the laser vector potential is given by $A(\tau)=a\cos \tau$, with the Lorentz-invariant phase $\tau=\omega(t-z)$ [see Eq.\,(\ref{A})]. After having left the laser field, the momentum components of Eq.\,(\ref{EoM}) become $P'_x = M$, $P'_y = 0$, $P'_z = M/2$. Here, the conditions $A(\tau_0)=a$ and $\xi=\xi_{\rm min}$ have been used. The Lorentz transformation to the lab frame now yields the momentum components of Eq.\,(\ref{typicalP}), with the reduced velocity of the frame transformation given by Eq.\,(\ref{beta}) below.

The assumptions that our simple man's model is based upon, can be infered from the saddle-point treatment of Sec.\,II.B. With regard to the electron-positron annihilation vertex, the saddle points $x_-$ are determined by
\begin{eqnarray}
\label{cosx}
\cos(\omega x_-) &=& -{u\over 8v} \pm \sqrt{{u^2\over 64v^2}+{n\over 4v}+{1\over 2}}.
\end{eqnarray}
Since $|u|\lesssim \xi\alpha m/\omega \ll |v|\approx \xi^2m/\omega$ and $n\approx 2v$ in our situation, Eq.\,(\ref{cosx}) becomes $\cos(\omega x_-)\approx \pm 1$. This means that the annihilation occurs at the laser phases $x_-$ when the amplitude of the vector potential (\ref{A}) attains a maximum. On the femtosecond time scale of the laser oscillation, the muon creation occurs directly afterwards, in agreement with one of our model assumptions. This circumstance moreover implies that, at $\xi=\xi_{\rm min}$, the muons are generated at rest in the c.m. frame, as we show now. The classical energy of a particle of charge $-e$ in a laser field given by Eq.\,(\ref{A}) reads
\begin{eqnarray*}
P_0(y_-) = P_0 - ea\omega{(\epsilon P)\over(kP)}\cos(\omega y_-) 
+ {e^2a^2\omega\over 2(kP)}\cos^2(\omega y_-)\,,
\end{eqnarray*}
where $P^\mu$ is the initial momentum and $y_-=t-z$. Consequently, the saddle point equation (\ref{SPE}) can be recast in the form
\begin{eqnarray}
P_+^0(y_-)+P_-^0(y_-) = q_0\,,
\end{eqnarray}
where $q_0\approx 2m\xi^2$ according to Eq.\,(\ref{q0}). We need to boost this 
equation into the c.m. frame. At maximum value of the laser vector 
potential, the reduced velocity of the electron (or positron) along the $z$ axis is
\begin{eqnarray}
\label{beta}
\beta = {p_z(x_-)\over p_0(x_-)}={\xi^2\over 1+\xi^2}\approx 1-{1\over\xi^2}\ ,
\end{eqnarray}
where $p_0(x_-)$ and $p_z(x_-)$ are the classical energy and momentum of the electron. This corresponds to a Lorentz factor of $\gamma \approx \xi/\sqrt{2}$ and yields
\begin{eqnarray*}
q_0' &=& \gamma(q_0-\beta q_z) 
 \approx \gamma\left[q_0-\left(1-{1\over 2\gamma^2}\right)(q_0-2p_0)\right] \nonumber\\
&\approx& {q_0\over 2\gamma}+2m\gamma \approx 2\sqrt{2}m\xi\,,
\end{eqnarray*}
where $q_z$ denotes the $z$-component of the virtual photon momentum .
At the threshold intensity [cf. Eq.\,(\ref{ximin})] we therefore obtain for the sum of the muon energies
\begin{eqnarray}
P_+^0(y_-)' + P_-^0(y_-)' = q_0' \approx 2\sqrt{2}m\xi_{\rm min} =2M\,,
\end{eqnarray}
i.e., the muons are created at rest.

\subsection{Analytical estimate of the total rate}
From Eq.\,(\ref{RPs}), one can derive by analytical means an approximate formula for the total rate of muon production from a single laser-driven Ps atom (see also \cite{muonPRD}). To this end, we consider the contribution to the rate stemming from the first term on the right-hand side of Eq.\,(\ref{uvw}):
\begin{eqnarray}
\label{tildeRPs}
\tilde R_{\rm Ps} &=& {\alpha^2\over (2\pi)^2}
{M^2\over 2^6 m^4\xi^8}
\int {d^3 Q_+\over Q_+^0} \int {d^3 Q_-\over Q_-^0} \sum_r \bar J_r^{\,2}\nonumber\\
&\times&\!\!\!\!\! \sum_{s_\pm,S_\pm}\!\!|u^\mu U_\mu|^2 \, \delta(q_+ + q_- - Q_+ - Q_- + rk).
\end{eqnarray}
Note that, when squaring Eq.\,(\ref{uvw}), eighty other terms of a similar structure
arise. According to our numerical calculations, the term accounted for in
Eq.\,(\ref{tildeRPs}) gives a major contribution to the total rate. 
With the help of the four-dimensional $\delta$ function, we can integrate over $d^3Q_+dQ_-^0$ and find
\begin{eqnarray}
\label{tildeRPs1}
\tilde R_{\rm Ps} &\approx& {\alpha^2\over (2\pi)^2}
{M^2M_\ast^2\over 2^8m^6\xi^8} \int d\phi_{Q_-} \int d\cos\theta_{Q_-} \nonumber\\
& & \times\sum_r \bar J_r^{\,2} \sum_{s_\pm,S_\pm}|u^\mu U_\mu|^2.
\end{eqnarray}
From the Jacobian of the energy-conserving $\delta$ function, a factor $\approx [2(1-\cos\theta_{Q_-})]^{-1}\approx (M_\ast/2m)^2$ arises here. We point out that this factor has been overlooked in \cite{muonPRD}, however, without affecting the main conclusions of this paper. The spin sum in Eq.\,(\ref{tildeRPs1}) can be converted into 
a product of two traces in the usual way,
\begin{eqnarray*}
\label{T}
\mathcal{T}_{uU} &:=& 
\sum_{s_\pm,S_\pm} \left| u^\mu U_\mu \right|^2 
 = {\rm Tr}\left( \gamma^\mu {\slash\!\!\!p_-+m\over 2m} 
\gamma^\nu {\slash\!\!\!p_+-m\over 2m} \right) \nonumber\\
& & \ \ \ \ \ \ \ \ \  \times
{\rm Tr}\left(\gamma_\mu {\slash\!\!\!\!P_-+M\over 2M} 
\gamma_\nu {\slash\!\!\!\!P_+-M\over 2M} \right).
\end{eqnarray*}
The standard trace technology yields
\begin{eqnarray*}
\label{Trace1}
\mathcal{T}_{uU} &=& 
{2\over m^2M^2}\left[ (p_-P_-)(p_+P_+) + (p_-P_+)(p_+P_-) \right] \nonumber\\
& & + {2(P_+P_-)\over M^2} + {2(p_+p_-)\over m^2} + 4.
\end{eqnarray*}
According to the kinematical analysis in Sec.\,III.A, a typical value of 
$\mathcal{T}_{uU}$ is 
\begin{eqnarray}
\label{T3}
\mathcal{T}_{uU} \approx {16 m^2\xi^4\over M^2},
\end{eqnarray}
which can be pulled out of the integration in Eq.\,(\ref{tildeRPs1}). 
We proceed by performing the further approximations
\begin{eqnarray}
\label{approx}
\int d\cos\theta_{Q_-} &\approx& {\theta_{\rm max}^2\over 2}\,,\
\int d\phi_{Q_-} \approx 4\phi_{\rm max}\,,\nonumber \\ 
\sum_r [J_r(U,\Delta V)]^2 &\approx& 1\,,\ \ \ \ \ \ \ \
\Delta V^{1/3} \approx |v_0|^{1/3}\,,   
\end{eqnarray}
where the maximum polar and azimuthal angles are given by Eqs.\,(\ref{thetamax})
and (\ref{phimax}). The first two estimates in Eq.\,(\ref{approx}) mean that the production rate is assumed as approximately constant within the mainly contributing angular domain. The factor of 4 in the $\phi_{Q_-}$ integral is due to the fact that
the integration range of interest extends over $(-\phi_{\rm max},\phi_{\rm max})$
and $(\pi-\phi_{\rm max},\pi+\phi_{\rm max})$. The third relation would exactly be true \cite{SFA3}, if the muon momenta and, thus, $U$ and $\Delta V$ were independent of $r$. The last estimate in Eq.\,(\ref{approx}) holds since $\Delta V=V-v_0$ and $v_0$ have a similar order of magnitude; i.e., in general $V$ and $v\approx v_0$ in Eq.\,(\ref{Vv}) compensate only partially, as is corroborated by our numerical calculations. As a result, we arrive at the formula
\begin{eqnarray}
\label{RPs2}
\tilde R_{\rm Ps} \approx {2^7\over\pi^2}{\alpha^2\over m^2\xi^2 a_0^3}
  \sqrt{1-{\xi_{\rm min}^2\over\xi^2}}
  {1\over\xi}\left({a_0\over\alpha\xi\lambda}\right)^{\!3}
  \left({m\over\omega\xi^4}\right)^{\!1/3},
\end{eqnarray}
which is the desired rate estimate for Ps decay into muons induced by a linearly polarized laser field. The analytical approximation (\ref{RPs2}) is confirmed by direct numerical evaluation of Eq.\,(\ref{RPs}), as shown in Fig.\,2 in Sec.\,IV. 

Equation\,(\ref{RPs2}) has an intuitive interpretation in terms of the cross 
section for the field-free process $e^+e^-\to\mu^+\mu^-$ \cite{Peskin}
\begin{eqnarray}
\label{sigma}
\sigma = {4\pi\over 3}{\alpha^2\over E_{\rm cm}^2}\sqrt{1-{4M^2\over E_{\rm cm}^2}}
\left(1+{2M^2\over E_{\rm cm}^2}\right)
\end{eqnarray}
and the electron-positron wave packet size, which is growing in the laser field due to 
quantum mechanical dispersion. In our case, the c.m. energy in Eq.\,(\ref{sigma}) 
is $E_{\rm cm}\approx 2\sqrt{2}m\xi$ according to Eq.\,(\ref{qsq}). We note that both Eq.\,(\ref{RPs2}) and (\ref{sigma}) vanish at the intensity (or energy) threshold.
The wave-packet spreading can be estimated as follows. In the (primed) electron-positron 
c.m. frame, the initial momentum spread is $\Delta p^{\prime}\sim 1/a_0$. The 
spreading during the recollision time $t_{\rm rec}^{\prime}$ can be estimated as 
\begin{eqnarray}
\label{spreading}
\Delta x^{\prime}\sim \Delta y^{\prime} \sim \Delta z^{\prime} \sim 
{\Delta p^{\prime}\over m}\,t_{\rm rec}^{\prime}.
\end{eqnarray}
Due to relativistic time dilation, the oscillation period in the c.m. frame
is largely enhanced so that the recollision time at $\xi \gg 1$ equals 
$t_{\rm rec}^{\prime}\sim \pi \gamma/\omega$, with the Lorentz factor 
$\gamma\sim \xi$ [see Eq.\,(\ref{beta})]. Consequently, at the collision the wave packet has spread to a size
$\Delta x^{\prime}\Delta y^{\prime}\Delta z^{\prime}\sim (\alpha\xi\lambda)^3$.
The corresponding particle current density leads to a reaction rate of
$R_{\rm Ps}^\prime \sim \sigma/(\alpha\xi\lambda)^3$. By boosting this rate
into the lab frame we obtain
\begin{eqnarray}
\label{RPs3}
R_{\rm Ps} \sim {\sigma\over\xi(\alpha\xi\lambda)^3}\ ,
\end{eqnarray}
which agrees with Eq.\,(\ref{RPs2}) up to factors of order unity.
This demonstrates that simple rate estimates for (nonresonant) nuclear or particle 
reactions in strong laser fields, which are based on semi-classical arguments and field-free cross sections, may be considered reliable (see, e.g., \cite{Corkum,Guido,collider}).

\subsection{Muon production from an $e^+e^-$ plasma}
So far, we have studied the laser-induced process ${\rm Ps}\to\mu^+\mu^-$. As was mentioned in the introduction, the corresponding reaction rate we denote as the {\it coherent} rate. It is given in approximate form by Eq.\,(\ref{RPs2}). When $N$ Ps atoms interact with the laser field, the total coherent rate $R_{\rm Ps}^{(N)}$ resulting from this ensemble increases proportionally with $N$, provided that each Ps atom independently creates a muon pair. The latter condition is fulfilled, when electrons and positrons originating from different Ps atoms do not interfere. This means that the electron wave packets from different atoms do not overlap, which is guaranteed according to the above, when $\alpha\xi\lambda\ll n^{-1/3}$, with the Ps density $n$.

We have seen in Sec.~III.C that the coherent rate is reduced by quantum mechanical wave-packet spreading of the initially well-localized electron-positron state. Spreading will be absent if, instead of forming a bound state, the electron and positron are initially free particles. In practice this could be realized by an $e^+e^-$ plasma. When driven by a strong laser field, $e^+e^-$ collisions occur inside the plasma, which can give rise to $\mu^+\mu^-$ production as before. The corresponding reaction rate we call the {\it incoherent} rate in order to distinguish it from the coherent rate resulting from Ps atoms. The incoherent rate scales with $N_+N_-$, where $N_+$ and $N_-$ are the numbers of electrons and positrons, respectively. In a neutral plasma ($N_+=N_-\equiv N$), the total incoherent rate $R_{e^+e^-}^{(N)}$ raises like $N^2$.

The incoherent rate is calculated via the square of the amplitude (\ref{See}). Integrals like the one in Eq.\,(\ref{int}) are absent then. Along the same lines followed in Sec.~III.C, one can derive an approximate expression for the incoherent rate of muon production from a nonrelativistic $e^+e^-$ plasma. It reads
\begin{eqnarray}
\label{Rfree}
R_{e^+e^-} \approx {1\over 2^3\pi^2}{\alpha^2\over m^2\xi^4}
\sqrt{1-{\xi_{\rm min}^2\over\xi^2}}\ {N_+N_-\over V},
\end{eqnarray}
with the interaction volume $V$. In our situation, the latter is determined by the laser focal spot size. 

Similar as before, Eq.\,(\ref{Rfree}) can be equipped with an intuitive meaning. To this end, we introduce the number $N_\mu$ of produced muons during the interaction time $T$ and rewrite Eq.\,(\ref{Rfree}) as
\begin{eqnarray}
\label{RfreeV}
\frac{R_{e^+e^-}}{V} = \frac{N_\mu}{VT} \sim \sigma \frac{n_+n_-}{\xi^2}.
\end{eqnarray}
Here, $\sigma$ is the field-free cross section of Eq.\,(\ref{sigma}) and $n_\pm=N_\pm/V$ are the electron and positron densities. The representation (\ref{RfreeV}) has the advantage that the combination $N_\mu/(VT)$ is Lorentz-invariant and can directly be transformed into the c.m. frame. The latter moves with the reduced velocity $\beta$ of Eq.\,(\ref{beta}) and the Lorentz factor $\gamma\sim\xi$. The particle densities $n_\pm'$ in the c.m. frame are related to the lab frame densities by $n_\pm=\gamma (n_\pm'-\beta j_z') = \gamma n_\pm'$ since the particle current density $j_z'$ vanishes. We can therefore recast Eq.\,(\ref{RfreeV}) into the form
\begin{eqnarray*}
\label{RfreeV2}
\frac{R_{e^+e^-}}{V} = \frac{R_{e^+e^-}^{\,\prime}}{V'} \sim \sigma n_+' n_-'
\end{eqnarray*}
or $R_{e^+e^-}^{\,\prime}\sim \sigma N_+ N_-/V'$. This means that in the c.m. frame, the number of events per volume and time is related to the corresponding cross section in the usual way.

We can compare the incoherent rate with the coherent rate for a single Ps atom by setting $N_+=N_-=1$ in Eq.\,(\ref{Rfree}). The interaction volume in the lab frame is determined by the wave-packet volume $V=(\alpha\xi\lambda)^3/\xi$ (see Sect. III.C), where a factor of $1/\xi$ accounts for the Lorentz contraction along the $z$ axis. With these choices, the incoherent rate gives the same order of magnitude as the coherent rate in Eq.\,(\ref{RPs3}).

To conclude this section we note that the coherent channel of muon production from Ps atoms evolves into the incoherent channel in the limit of large wave packet size ($\alpha\xi\lambda\gtrsim n^{-1/3}$). On the one hand, with increasing wave packet size the coherent rate is more and more suppressed; on the other hand, electrons (positrons) stemming from different Ps atoms start to overlap and the gas of Ps atoms transforms into an $e^+e^-$ plasma. Therefore, in the limit of large wave-packet size, the incoherent reaction rate (\ref{Rfree}) will eventually dominate the muon production process.

\subsection{The influence of the laser polarization}
In a previous paper, we have treated the process ${\rm Ps}\to\mu^+\mu^-$ in a laser field of circular polarization \cite{muonPRD}. The threshold intensity in this case is larger by a factor of two, since the same minimum peak field strength is needed. From the investigation of laser-driven recollision processes in atoms or molecules (e.g., high-harmonic generation) it is known that the efficiency of these processes is strongly suppressed in a circularly polarized laser field \cite{Becker,HHG,HHG2}. The reason is that the ionized electron is forced onto a circular trajectory and does not return to the parent ion, which stays at rest due to its heavy mass. In our case, the situation is different as both binding partners move along circular orbits (within the polarization plane) under the influence of a circularly polarized field. But still, we also find a strong reduction of the reaction ${\rm Ps}\to\mu^+\mu^-$ when the laser polarization is circular. The corresponding rate approximately amounts to
\begin{eqnarray}
\label{RPscirc}
\tilde R_{\rm Ps}^{(c)} \approx \frac{\sigma}{a_0^3} \left(\frac{a_0}{\lambda\xi_c}\right)^4 \left( {4m\over\omega\xi_c} \right)^{2/3},
\end{eqnarray}
with $\xi_c=ea/m$. Here, the Volkov state normalization of Eq.\,(\ref{Vol}) and the approximate treatment of the $\delta$-function in Eq.\,(\ref{tildeRPs1}) has been applied. We emphasize that this leads to a correction factor of $\xi_c^{-2}$ as compared to Eq.\,(52) in \cite{muonPRD}, resulting in an even smaller production probability than predicted there. The physical conclusions of this paper thus remain unchanged. In particular, the rate suppression in a circularly polarized field is mainly expressed by the damping factor $[a_0/(\lambda\xi_c)]^4\sim 10^{-25}$ in Eq.\,(\ref{RPscirc}). This factor can be attributed to the classical motion of the electron and positron, which corotate on opposite sides of a circle of radius $\sim\lambda\xi_c$ in the polarization plane. The interparticle distance is thus larger than their wave-packet extensions. This is in contrast to the case of a linearly polarized field, where the classical trajectories periodically meet and the rate damping results from quantum mechanical dispersion (cf. Sec.~III.C). Equation~(\ref{RPscirc}) can also be written in the form
\begin{eqnarray}
\label{RPscirc2}
{\tilde R}_{\rm Ps}^{(c)} \approx {15\over 16} {\alpha^2\over m^2\xi_c^2} 
{1\over\xi_c\alpha(\xi_c\lambda)^3} \left({\omega\over m\xi_c^2}\right)^{1/3},
\end{eqnarray}
which is more similar to Eq.\,(\ref{RPs2}) and allows an interpretation along the lines of Eqs.\,(\ref{sigma})-(\ref{RPs3}). According to the above, the electron-positron distance in the c.m. frame amounts to
$\Delta x^{\prime}\sim \Delta y^{\prime} \sim \lambda\xi_c\,,
\Delta z^{\prime} \sim \alpha\xi_c\lambda$ [cp. Eq.\,(\ref{spreading})]. Hence, as before in Eq.\,(\ref{RPs3}), the total reaction rate can be expressed as
$\tilde R_{\rm Ps}^{(c)} \sim
\sigma/(\xi_c\Delta x^{\prime}\Delta y^{\prime}\Delta z^{\prime})$,
where we have dropped factors which are of order unity at $\xi_c\gtrsim M/m$.

The incoherent rate of muon production from a nonrelativistic $e^+e^-$ plasma driven by a circularly polarized laser field reads 
\begin{eqnarray}
\label{RfreeCirc}
R_{e^+e^-}^{(c)} \approx\frac{9}{16\pi}\frac{\alpha^2}{m^2\xi_c^4}\frac{N_+N_-}{V}\,.
\end{eqnarray}
The formula involves an additional factor of $\approx\xi_c^{-2}$ in comparison with Eq.\,(59) in \cite{muonPRD}, as well.
Since quantum wave-packet spreading and classical motion are not crucial here, the incoherent rate is of the same order of magnitude as the corresponding one in a laser field of linear polarization [cf. Eq.\,(\ref{Rfree})]. The incoherent muon yield in a circularly polarized field is slightly larger, which can be attributed to the fact that the field amplitude is constant and not oscillating, as it is for a lineraly polarized wave. Similar results are known from other strong-field phenomena such as $e^+e^-$ pair creation in combined laser and Coulomb fields (see, e.g., \cite{Milstein}).

\section{Numerical results and Discussion}

\subsection{Total muon yields}
On the basis of Eq.\,(\ref{RPs}) we have also performed numerical calculations of the laser-driven process ${\rm Ps}\to\mu^+\mu^-$. The high-order generalized Bessel functions were evaluated with the asymptotic expansion (\ref{genbesasy}). Figure~2 shows the dependence of the total reaction rate on the laser peak field strength for a near-infrared laser frequency of $\omega = 1$\,eV and a single Ps atom. For comparison, the figure also contains the analytical rate estimate in Eq.\,(\ref{RPs2}) and the corresponding results for a circularly polarized laser field. For both polarization states, the agreement between the numerical simulation and the approximate expression is reasonably good. The reaction rates in the case of circular polarization are about eight orders of magnitude smaller than in the case of linear polarization. The very pronounced impact of the laser polarization on the process efficiency was discussed in Sec.~III.E.

In a linearly polarized driving field, the muon production rate starts from zero at the energetic threshold [see Eq.\,(\ref{RPs2})] and reaches maximum values of the order of 10$^{-10}$ s$^{-1}$ at laser intensities of about $5\times 10^{22}$ W/cm$^2$ ($\xi\approx 170$). The overall behaviour of the rate is similar to that of the cross section for the field-free process $e^+e^-\to\mu^+\mu^-$ in Eq.\,(\ref{sigma}). At higher intensities, however, the rate decreases more strongly than the cross section, since the decrease results not only from the typical high-energy dependence $1/E_{\rm cm}^2 $ but also from the enhanced wave-packet spreading [cf. Eq.\,(\ref{RPs3})].

\begin{figure}[h]
\begin{center}
\resizebox{8cm}{!}{\includegraphics{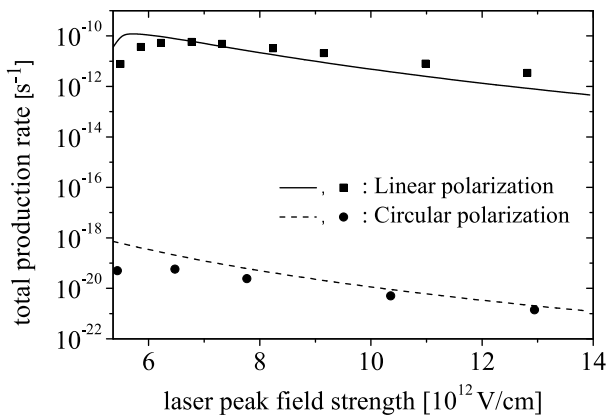}}
\caption{Total rates for the process ${\rm Ps}\to\mu^+\mu^-$ induced by an 
intense near-infrared laser field ($\omega = 1$\,eV), as a function of the laser peak field strength. The black squares and the solid line refer to a linearly polarized laser field and show the results of numerical calculations based on Eq.\,(\ref{RPs}) and the analytical estimate in Eq.\,(\ref{RPs2}), respectively. The black circles and the dashed line show the corresponding results for a laser field of circular polarization.} 
\end{center} 
\end{figure}

For comparison we note that the rate for muon creation in a hypothetical large-scale $e^+e^-$ collider experiment, that employs a single electron and a single positron only, would be of the order of $10^{-20}$\,s$^{-1}$ \cite{muonPRD}. The rate resulting from the ``$e^+e^-$ micro-collider'' based on Ps is orders of magnitude larger, due to the small impact parameters and the correspondingly large current densities (``luminosities'') achievable with linearly polarized driving fields. A real collider experiment, however, employs beams of $\sim 10^{10}$ particles and reaches reaction rates of about 1\,s$^{-1}$ this way. This number, in turn, is orders of magnitude larger than the coherent rate from a single Ps atom. In Sec.~IV.C below, we discuss the implications of these rate estimates for an experimental realization of the laser-induced process ${\rm Ps}\to\mu^+\mu^-$.

\subsection{Muon energy distribution}

In Fig.\,3, partial production rates with respect to the number $r$ of 
absorbed laser photons are shown (i.e., the contributions to the total rate 
stemming from a net absorption of $r$ laser photons in the production process). 
The photon number is given in units of $r_0 \equiv m\xi^2/(2\omega)$.
Since $r\omega\approx P_+^0+P_-^0$ [see Eq.~(\ref{typicalP})], the partial rates reflect the energy distribution of the created muons. The distribution is broad, indicating a substantial width of the energy which is available in the process and transformed into muons.

First we emphasize that the broadening of the muon energy spectra is not due to the initial momentum spread in the Ps atom, which amounts to $\Delta p\sim m\alpha$. 
In accordance with Eq.\,(\ref{sigma}), the c.m. energy associated with the $e^+e^-$ annihilation is $E_{\rm cm}=[(q_+ + q_- -nk)^2]^{1/2}\sim 200\,$MeV. It depends on the initial momentum $p$ via the effective particle momenta $q_\pm$ [see Eq.\,(\ref{q})]. The initial momentum spread $\Delta p$ translates into a longitudinal width $\Delta (q_+ + q_-)_z\sim m\alpha^2\xi^2$, whereas the transversal momentum components add up to zero. Since $\Delta p$ is small, the number $n$ of photons absorbed at the electron-positron vertex of the Feynman graph varies within narrow bounds [see Eq.\,(\ref{qsq})]. As a result, the width of the collision energy $\Delta E_{\rm cm}\sim E_{\rm cm}\alpha^2\sim 10\,$keV is small.

The main limitation for a well-defined effective collision energy comes from the fact that the total number $r$ of absorbed laser photons is not fixed, but has a large width $\Delta r$ (see Fig. 3). This is because the number of photons absorbed at the muon-antimuon vertex varies within a wide range. The contribution of these laser photons leads to an enhancement of the available energy, that can be transformed into muons. The effective collision energy is accordingly  given by ${\tilde E}_{\rm c.m.}=[(q_+ +q_- + rk)^2]^{1/2}>E_{\rm cm}$, as indicated by the $\delta$ function in Eq.~(\ref{Ssq}). The available energy must be larger than $E_{\rm cm}\approx 2M$, because the muons have to be produced with their effective mass $M_\ast>M$ in the laser field. The energetic difference $\Delta M^2 = 2(M_\ast^2-M^2)$ is supplied by the minimal number of laser photons $r_{\rm min}$, according to Eq.\,(\ref{rmin}). 
At the laser intensity threshold (where the process probability vanishes), only photon numbers $r\approx r_{\rm min}$ contribute and the muons are produced with a sharp energy $P_0\approx M^2/m\approx 22$\,GeV. We have seen in Sec.\,III.B that in this case the muons are created in the field at rest in the c.m. frame.
For larger intensities, the range of contributing photon numbers quickly increases and attains a typical size of $\Delta r\sim 10^{10}$ for the parameters considered in Fig.~3.
This leads to a corresponding spread $\Delta r\omega\sim 10\,$GeV of the muon energies 
in the laboratory frame (or $\Delta r\omega'\sim 100\,$MeV in the c.m. frame).
The longitudinal and transversal momentum widths amount to $\Delta P_z \sim \Delta r\omega \sim 10\,$GeV and $\Delta P_\perp \sim \Delta r\omega' \sim 100\,$MeV, where the first estimate results from the relation $P_z\approx P_0$, and the second from $U\approx r$, in accordance with the properties of the Bessel functions. 

The shape of the curves in Fig.\,3 can be understood in more detail via the kinematical analysis in Sec.\,III.\,A. First, according to Eq.\,(\ref{rmin}), the minimal number of laser photons required from kinematical constraints amounts to $r_{\rm min} = 2.2\times 10^{10}$, independent of the laser intensity. If we express this number 
with respect to the respective values of $r_0$, we find $r_{\rm min}/r_0 = 3.0$ 
($\xi = 170$), $2.2$ ($\xi = 200$), and $1.4$ ($\xi = 250$), as is displayed in 
Fig.\,3. The partial reaction rate for $r = r_{\rm min}$ is always zero. Furthermore,  the curves exhibit maxima at $r\approx \bar r$, in agreement with Eq.\,(\ref{barr}); the typical photon number $\bar r$ is increasing with the applied laser intensity. We point out that the width of the curves is significantly
larger than in the case of circular laser polarization (see Fig.\,3 and Ref.\,\cite{muonPRD}). This can be understood by observing that, in a linearly polarized field, $e^+e^-$ collisions not only occur at the optimum laser phase (which is predicted by the simple man's model and gives rise to the absorption of
$\bar r$ photons), but also at earlier or later phases. Hence the collision energy 
varies, which broadens the energy distribution of the created particles. Different from that, in the circular polarization case all creation phases are equivalent, so that the width of the partial production rate is purely of quantum mechanical origin.

\begin{figure}[h]
\begin{center}
\resizebox{8cm}{!}{\includegraphics{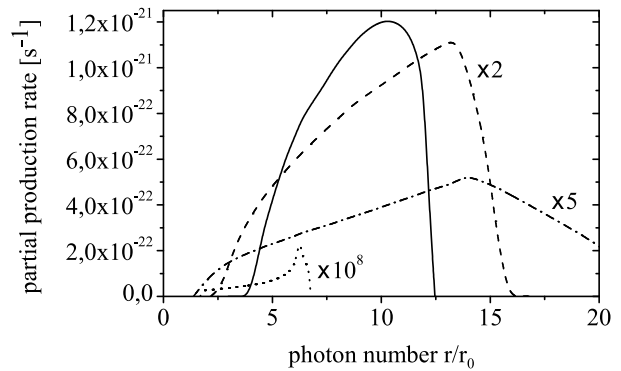}}
\caption{Partial rates for the laser-driven decay $\mbox{Ps}\to \mu^+\mu^-$, as a function of the number of absorbed laser photons $r$. The linearly polarized laser field has a frequency of $\omega = 1$\,eV and an intensity parameter of $\xi = 170$ (solid line), 200 (dashed line, enhanced by a factor of 2), and 250 (dash-dotted line, enhanced by a factor of 5), respectively. For comparison, the dotted line shows the corresponding result for a circularly polarized field with $\xi_c=250$; it is enhanced by a factor of $10^8$. The photon number is given in units of $r_0 = m\xi^2/(2\omega)$.} 
\end{center} 
\end{figure}

\subsection{Experimental feasibility}
The maximum coherent rate resulting from a single laser-driven Ps atom is of the order of $R_{\rm Ps}\sim 10^{-10}$\,s$^{-1}$ (see Fig.\,2). The total process probability is obtained by multiplying the rate with the interaction time, which is given by the laser pulse duration. The high intensities required to ignite the process ($I\sim 10^{22}$--$10^{23}$ W/cm$^2$) are attained in short laser pulses, with typical durations in the ps--fs domain. The muon yield per laser pulse from a single Ps atom is therefore very small ($\lesssim 10^{-20}$). 

Since the total probability of the coherent process increases linearly with the number of Ps atoms involved, high Ps densities are desirable. In recent years, a remarkable progress has been made in the efficient production, accumulation and trapping of Ps atoms \cite{Surko}. The highest Ps density achieved so far is of the order of $n=10^{15}$\,cm$^{-3}$ \cite{BEC}. At this density, a typical laser focal volume of $V=(10\lambda)^3\sim 10^{-9}$\,cm$^3$ contains a million atoms. The total coherent reaction rate could accordingly be increased by this number to $R_{\rm Ps}^{(N)}\sim 10^{-4}$\,s$^{-1}$. There are proposals how to generate even denser Ps samples ($n\gtrsim 10^{18}$\,cm$^{-3}$) with regard to the formation of a Ps Bose-Einstein condensate \cite{BEC} or experimental studies on antihydrogen \cite{antihydrogen}. Such high densities, however, violate the condition $\alpha\xi\lambda\ll n^{-1/3}$ for coherent collisions, so that electron (positron) wave packets from different Ps atoms will considerably overlap at the collision. The muon creation then proceeds via the incoherent channel. 
Incoherent muon production from a laser-driven $e^+e^-$ plasma requires high target densities, as well, because of the small focal spot size of strong laser pulses, which determines the interaction volume. By assuming again a typical focal volume of $V=(10\lambda)^3\approx 10^{-9}$\,cm$^3$, which contains $N_\pm=10^7$ particles at a  plasma density of $10^{16}$\,cm$^{-3}$ \cite{Surko,antihydrogen}, we obtain a total incoherent rate of $R_{e^+e^-}^{(N)}\sim 10^{-2}$\,s$^{-1}$ at $\xi\approx 230$ [see Eq.\,(\ref{Rfree})]. In view of the short duration of strong laser pulses and the above  values of $R_{\rm Ps}^{(N)}$ and $R_{e^+e^-}^{(N)}$, both the coherent and incoherent channels of laser-driven muon pair production seem to be hardly observable in experiment
at first sight.

The wave packet spreading, however, which has a detrimental impact on the coherent reaction rate, can be strongly reduced by applying more complex driving-field geometries than a single laser wave. A suitable configuration is formed by two counterpropagating laser laser waves, as is realized for example by the Astra Gemini system at the Rutherford Appleton Laboratory or the JETI ``photon collider'' at the University of Jena \cite{RAL}. The resulting Ps dynamics has been investigated in Ref.~\cite{collider} by means of a classical Monte-Carlo simulation for circularly polarized laser waves of equal frequency and intensity. It was found that the wave packet spreading is largely reduced since the recollision times are much shorter than in a single laser wave. Moreover, the resulting laser magnetic field is oriented along the electric field and induces a focussing force \cite{Corkum}. As a consequence, the particle wave packets are as small as (10\AA)$^3$ at the recollision. This implies  that the muon yield could be increased by $\sim 10$ orders of magnitude as compared to the case of a single driving laser field [see Eq.~(\ref{RPs3})]. Moreover, due to the small wave packet size, higher Ps densities can be exploited, without violating the condition for coherent collisions. At the envisaged density of $n\gtrsim 10^{18}$\,cm$^{-3}$ \cite{BEC,antihydrogen}, a total coherent reaction rate of $R_{\rm Ps}^{(N)}\sim 10^7$\,s$^{-1}$ could be achieved, with $N=10^9$. Under these circumstances, the observation of coherent muon pair production from laser-driven Ps becomes feasible at high laser repetition frequency $\sim$Hz-kHz \cite{Bahk,ELI}. The resulting number of events approaches 1\,s$^{-1}$, which is comparable to the event rates at large-scale accelerators (see Sec.\,IV.A). 
As to the kinematics of the created particles, we note that the laboratory and c.m. frames coincide here due to the symmetric setup. The expected emission pattern is therefore approximately isotropic, in contrast to the muon kinematics resulting in the case of a single driving laser field (see Sec.\,III.A).

We point out that also the incoherent production channel is enhanced in the counterpropagating beam-geometry, since the c.m. motion is absent. In accordance with Eq.~(\ref{RfreeV}), the enhancement factor amounts to $\xi^2\sim 10^4$--10$^5$.
Another promising field combination is formed by a single laser wave and an additional static magnetic field, which is spatially confined \cite{collider}. This geometry allows for higher collision energies, as the large longitudinal momentum ($q_z\sim m\xi^2$) of the electron and positron is transversally redirected by the magnetic kick and can thus be exploited for particle production.

To conclude this section, we point out the following technical aspect. Our calculation assumes that the electron and positron trajectories lie completely within the laser field. The transverse extent of the classical trajectories along the laser electric-field axis is $\Delta x\sim \lambda\xi\sim 10^2\lambda$. This is larger than covered by a typical laser focal diameter. On the other hand, there is practically no motion along the magnetic-field direction, apart from wave packet spreading ($\Delta y\sim\alpha\xi\lambda\approx\lambda$). In order not to waste laser power, it would therefore be useful to employ laser foci of elliptical shape: prolonged along the polarization axis and compressed in perpendicular direction. This way, a laser focal area of order $\Delta x\Delta y\sim (10\lambda)^2$ could fully accomodate the electron and positron motion in the field.

\section{Summary and Conclusion}
The creation of a muon pair by $e^+e^-$ annihilation in an intense laser wave of linear polarization was calculated. The initial electrons and positrons were assumed to form a gas of Ps atoms or a nonrelativistic plasma, whereas the energy required for the process is provided by the laser field. The minimum field intensity amounts to a few $10^{22}$ W/cm$^2$ in the near-infrared frequency domain, corresponding to an intensity parameter of $\xi\approx 150$. These laser parameters are within experimental reach \cite{Bahk,ELI}. 

Our calculation proceeded within the framework of laser-dressed QED, employing the strong-field approximation. The initial bound state was expanded into a superposition of products of Volkov states, weighted by the Compton profile of the Ps ground state. The produced muons were also described by relativistic Volkov states. The process amplitude has been evaluated in two ways. First, the standard method of expansion into Fourier series was applied. The appearance of generalized Bessel functions of very high order $\sim 10^{11}$ represents a technical difficulty here. In a second approach, we evaluated the amplitude by the saddle-point integration method and obtained this way a suitable asymptotic expansion for the generalized Bessel functions, which has been used for numerical computations. The numerical results on the total reaction rates are in good agreement with corresponding analytical estimates. The latter allow for an intuitive understanding of the process in terms of the field-free muon production cross section and the size of the colliding electron and positron wave-packets.

The produced muons are characterized by ultrarelativistic energies and very narrow emission angles around the laser propagation direction. The muonic life time is accordingly enhanced by relativistic time dilation. The peculiar muon kinematics is explicable in terms of a semi-classical simple man's model. It essentially arises from the fact, that the electron and positron are forced by the laser field to longitudinal velocities close to the speed of light, i.e., the c.m. frame is moving at high velocity with respect to the laboratory frame. The energy distribution of the muons is broad, because the number of photons absorbed  from the laser field during the production process varies over a wide range. 

As to the total reaction rates, we found that a high scattering luminosity is achieved in the case with initial Ps atoms. This is due to coherent $e^+e^-$ collisions at small impact parameters, which result from the microscopic extension of the initial state and the symmetric electron-positron dynamics in the laser field. Limitations on the process efficiency are set by quantum mechanical wave-packet spreading which dilutes the particle density at the collision. The resulting coherent reaction rate from a dense Ps sample is about 10$^{-4}$\,s$^{-1}$ for a laser of intensity $\approx 5\times 10^{22}$ W/cm$^2$ and focal volume $\approx(10\lambda)^3$. This value is several orders of magnitude larger than the corresponding reaction rate in a circularly polarized field, where the process is suppressed by the classical electron and positron trajectories. The incoherent rate for muon production from a laser-driven $e^+e^-$ plasma is essentially insensitive to the field polarization and reaches values of $\sim 10^{-2}$\,s$^{-1}$ at high plasma densities.

The field-induced process ${\rm Ps}\to\mu^+\mu^-$ is experimentally accessible in a setup that employs a high-density Ps target ($n\sim 10^{18}$\,cm$^{-3}$) and counterpropagating laser beams to control the wave-packet spreading. The coherent production rate could be increased to $\sim 10^7$\,s$^{-1}$ this way, which translates into an effective muon yield of about $1$\,s$^{-1}$ when the laser pulse duration ($\sim 100$\,fs) and repetition frequency ($\sim\,$Hz-kHz) are taken into account. The event rate achievable with the laser-driven ``$e^+e^-$ micro-collider'' in the crossed-beams configuration is similar to those at large-scale accelerator facilities. An experimental observation of laser-induced muon pair creation from Ps decay, though challenging, will therefore be rendered feasible by the next-generation of high-power laser devices.

\vspace{1cm}

\renewcommand{\theequation}{A\arabic{equation}}
\setcounter{equation}{0}
\appendix
\begin{center}
  {\bf APPENDIX A}
\end{center}
In the derivation of Eq.\,(\ref{Ssq}) in Sec.\,II.A, we neglected weak dependences on the bound-state momentum $\mbox{\boldmath$p$}$ in the 
transition amplitude when performing the integral over the Ps momentum 
distribution. As regards kinematic terms appearing as linear factors, this
approximation is certainly justified. But we also ignored the 
$\mbox{\boldmath$p$}$ dependence in the argument of the $\delta$ function in 
Eq.\,(\ref{Ssq}), which is a more subtle simplification. In this appendix we 
show that this approximation is indeed justified.

With the $\delta$ function included, the integral in Eq.\,(\ref{barJr}) reads
\begin{eqnarray}
\label{tildeJr}
\tilde J_r &=& \int {d^3p\over (2\pi)^3}\,\tilde\Phi(\mbox{\boldmath$p$}) 
J_r(U-u,V-v) \nonumber\\
& & \times\,\delta(q_+ + q_- - Q_+ - Q_- +rk).
\end{eqnarray}
According to Eq.\,(\ref{q}), the effective momenta are weakly dependent on 
$\mbox{\boldmath$p$}$ (see also Sec.\,IV.B). We perform a Taylor expansion up to leading order
\begin{eqnarray*}
q_+^\mu + q_-^\mu \approx q_+^\mu(0) + q_-^\mu(0) 
                   + \Delta q(\mbox{\boldmath$p$}){k^\mu\over\omega}\ ,
\end{eqnarray*}
where $q_\pm^\mu(0)$ denote the values of $q_\pm^\mu$ at $\mbox{\boldmath$p$}=0$ 
and
\begin{eqnarray}
\label{deltaq}
\Delta q(\mbox{\boldmath$p$}) = {(p_z^2-p_\perp^2)\xi^2\over 2m}.
\end{eqnarray}
Within the momentum range given by $\tilde\Phi(\mbox{\boldmath$p$})$, the
correction term can be as large as 
$\Delta q(\mbox{\boldmath$p$})\sim m\alpha^2\xi^2/8$. 
Since it is multiplied by the wave four-vector in Eq.\,(\ref{tildeJr}), we can
consider $\Delta q(\mbox{\boldmath$p$})/\omega\sim 10^5$ as a correction to the 
photon number $r$. Recall that a typical value is $r\sim 10^{11}$ so that 
$r^{1/3}\sim 10^4$. Hence, by the Bessel function properties, the correction 
to the photon number induced by the term (\ref{deltaq}) could, in general, be 
important.

By writing the square of the amplitude as a product of two terms, we can express
the reaction rate as
\begin{eqnarray}
\label{Rsim}
R_{\rm Ps}\!\! &\sim&\!\!\! \int{d^3Q_+\over Q_+^0}\int{d^3Q_-\over Q_-^0} \sum_{r,r'}
\int\! d^3p\,\tilde\Phi(\mbox{\boldmath$p$}) J_r(U-u,V-v) \nonumber\\
&\times&
 \delta\left(q_+(0)+q_-(0)+\Delta q(\mbox{\boldmath$p$})k/\omega-Q_+-Q_-+rk\right)
 \nonumber\\
&\times& \int d^3p'\,\tilde\Phi(\mbox{\boldmath$p$}') J_r(U-u',V-v') \nonumber\\
&\times& 
\delta\left(q_+(0)+q_-(0)+\Delta q(\mbox{\boldmath$p$}')k/\omega-Q_+-Q_-+r'k\right)
\nonumber\\
\,
\end{eqnarray}
where linear kinematic factors were suppressed in the notation.
Both $\delta$ functions in Eq.\,(\ref{Rsim}) contain the term 
$\delta(Q_+^\perp + Q_-^\perp)$. One of these we exploit to integrate by 
$Q_+^\perp$, and the integral over $Q_{+,z}$ is performed with the help of 
$\delta(q_{+,z}(0)+q_{-,z}(0)+\Delta q(\mbox{\boldmath$p$}')-Q_{+,z}-Q_{-,z}+r'\omega)$.
The latter $\delta$ function fixes the value of $Q_{+,z}$ to
\begin{eqnarray*}
\label{Qz}
\tilde Q_{+,z} &\equiv& 
q_{+,z}(0)+q_{-,z}(0)+\Delta q(\mbox{\boldmath$p$}') - Q_{-,z} + r'\omega.
\end{eqnarray*}
Another two of the remaining $\delta$ functions coincide and we obtain
(with the interaction length $L$) 
\begin{widetext}
\begin{eqnarray}
R_{\rm Ps} &\sim& L^3 \int {d^3 Q_-\over Q_-^0} \sum_{r,r'}
\int d^3p \,\tilde\Phi(\mbox{\boldmath$p$})  J_r(U-u,V-v)
\delta\left(q_+^0(0) + q_-^0(0) +\Delta q(\mbox{\boldmath$p$}) 
- \tilde Q_+^0 - Q_-^0 +r\omega\right) \nonumber\\
& &\times \int {d^3p'\over\tilde Q_+^0} \tilde\Phi(\mbox{\boldmath$p$}') 
J_r(U-u',V-v') \delta\left(q_+^0(0) + q_-^0(0) +\Delta q(\mbox{\boldmath$p$}') 
- \tilde Q_+^0 - Q_-^0 +r'\omega\right) \nonumber\\
&=& L^3 \int {d^3 Q_-\over Q_-^0}\left|\sum_r
\int {d^3p\over\sqrt{\tilde Q_+^0}}\,\tilde\Phi(\mbox{\boldmath$p$}) J_r(U-u,V-v) 
\delta\left(q_+^0(0) + q_-^0(0)+ \Delta q(\mbox{\boldmath$p$}) - \tilde Q_+^0 - Q_-^0 +r\omega\right)\right|^2,
\end{eqnarray}
\end{widetext}
where in the last step we made use of the fact that $\tilde Q_{+,z}$ and thus 
$\tilde Q_+^0$ can also be expressed in terms of $\mbox{\boldmath$p$}$ and $r$, since 
$\Delta q(\mbox{\boldmath$p$})+r\omega = \Delta q(\mbox{\boldmath$p$}')+r'\omega$.
By performing a Taylor expansion 
\begin{eqnarray*}
\tilde Q_+^0 = \sqrt{M_\ast^2+Q_\perp^2+Q_{+,z}^2}
\approx Q_+^0(0) + \Delta q(\mbox{\boldmath$p$})\,,
\end{eqnarray*}
we observe that the leading $\mbox{\boldmath$p$}$ dependence drops out of the 
argument of the $\delta$ function. Hence, we arrive at the desired result
\begin{eqnarray*}
R_{\rm Ps}\! &\sim&\! L^4\!\! \int {d^3 Q_-\over Q_-^0 Q_+^0(0)} \sum_r
\left|\int d^3p\,\tilde\Phi(\mbox{\boldmath$p$}) J_r(U-u,V-v) \right|^2\nonumber\\
& & \times\,\delta\left(q_+^0(0) + q_-^0(0) - Q_+^0 - Q_-^0 +r\omega\right),
\end{eqnarray*}
which shows that the argument of the $\delta$ function in Eq.\,(\ref{Ssq}) may
be taken at $\mbox{\boldmath$p$}=0$.

\vspace{1cm}

\renewcommand{\theequation}{B\arabic{equation}}
\setcounter{equation}{0}
\appendix
\begin{center}
  {\bf APPENDIX B}
\end{center}
In this appendix we calculate the integral in Eq.\,(\ref{int}):
\begin{eqnarray}
\label{Jr}
 \bar{J_r} = \int{d^3p\over (2\pi)^3}\,\tilde\Phi(\mbox{\boldmath$p$})J_r(U-u,V-v)\,,
\end{eqnarray}
where
\begin{eqnarray}
\label{arg2}
u \approx {2\sqrt{2}\xi p_x\over\omega}\,,\ \ 
v \approx -{m\xi^2\over 2\omega} \left(1+{p_z^2-p_\perp^2\over 2m^2}\right)
\end{eqnarray}
according to Eq.\,(\ref{arguv}).
In the Taylor expansions on the right-hand side of these equations we kept those 
terms which are of importance according to the properties of the function 
$J_r(U-u,V-v)$. By virtue of Eq.\,(\ref{genbes}) we can expand Eq.\,(\ref{Jr}) 
into series $\bar J_r = \sum_s \bar J_{r,s}$ with
\begin{eqnarray}
\label{Jrs}
\bar J_{r,s} = \int{d^3p\over (2\pi)^3}\,\tilde\Phi(\mbox{\boldmath$p$})
J_{r-2s}(U-u)J_s(V-v).
\end{eqnarray}
From the kinematical analysis in Sec.\,III.A we know that $V\approx v$. As a 
consequence, the magnitude of the argument $V-v$ is much smaller than the
order $r$ of the generalized Bessel function. This is corroborated by our 
numerical calculations, according to which $|V-v| $ is smaller than 
$r$ by an order of magnitude in the integration region of interest.
In order to obtain significant values of $\bar J_{r,s}$ it is required that 
$s\approx V-v$ and $r-2s\approx U-u$.
Since $s\ll r$ and $u\sim\xi p_x/\omega\lesssim \xi\alpha m/\omega\ll U\sim \xi^2 m/\omega$, this implies $r\approx U$. Now, $J_{r-2s}(U-u)$
is almost constant for $|u|\lesssim u_{\rm max} \sim r^{1/3}$ and strongly
damped otherwise. Therefore, we can perform the $p_x$ integration
in Eq.\,(\ref{Jrs}) approximately and obtain
\begin{eqnarray}
\label{Jrsx}
\bar J_{r,s} \approx 2p_x^{\rm max}J_{r-2s}(U)
\int{dp_ydp_z\over (2\pi)^3}\,\tilde\Phi(\mbox{\boldmath$p$})J_s(V-v)\,,
\end{eqnarray}
where
\begin{eqnarray*}
p_x^{\rm max} \equiv {\omega\over 2\sqrt{2}\xi}\left({m\xi^2\over\omega}\right)^{1/3}
\end{eqnarray*}
is the $p_x$ value corresponding to $u_{\rm max}$ by Eq.\,(\ref{arg2}), and 
$\tilde\Phi(\mbox{\boldmath$p$})$ may be taken at $p_x=0$ since 
$p_x^{\rm max}\ll m\alpha$. By introducing the variables $\sigma\equiv p_y^2+p_z^2$ 
and $\tau\equiv p_y^2-p_z^2$, Eq.\,(\ref{Jrsx}) becomes
\begin{eqnarray*}
\bar J_{r,s} \approx {2p_x^{\rm max}\over(2\pi)^3}\,J_{r-2s}(U)
\int_0^\infty d\sigma \tilde\Phi(\sigma)I_s(\sigma)\,,
\end{eqnarray*}
with 
\begin{eqnarray}
\label{Isigma}
I_s(\sigma) &\equiv& \int_{-\sigma}^{+\sigma}{d\tau\over\sqrt{\sigma^2-\tau^2}}
\,J_s\left[V-v_0\left(1-{\tau\over 2m^2}\right)\right] \nonumber\\
&=& \int_{\Delta V-\zeta}^{\Delta V+\zeta} 
{J_s(x)dx\over\sqrt{\zeta^2-(x-\Delta V)^2}}\ ,
\end{eqnarray}
where the substitution $x\equiv V-v_0[1-\tau/(2m^2)]$ was made in the second step 
and the definitions $\zeta\equiv -\sigma v_0/(2m^2)$, $\Delta V\equiv V-v_0$ were employed.
In Eq.\,(\ref{Isigma}) the width of the function $J_s(x)$ is determined by
$s^{1/3}\sim \Delta V^{1/3}$, whereas the width of the square root in the
denominator is given by $\zeta$. In order to evaluate the integral $I_s(\sigma)$,
we accordingly divide the integration range into regions and show
\begin{eqnarray}
\label{Isigma2}
 I_s(\sigma) \approx \left\{\begin{array}{l}
   \displaystyle{ \pi J_s(\Delta V)
    \ \ \ \ (\zeta \ll \Delta V^{1/3}),}\\
   \\
   \displaystyle{ {1\over \zeta}
   \ \ \ \ \ \ \ \ \ \ \ \ \ \ (\zeta\gg\Delta V^{1/3}\ {\rm and}\ |s-\Delta V|\ll\zeta),}\\
   \\
   \displaystyle{ {1\over\sqrt{\zeta\Delta V^{1/3}}}
   \ \ \ (\zeta\gg\Delta V^{1/3}\ {\rm and}\ |s-\Delta V|\sim\zeta).}   
                 \end{array}\right.\nonumber\\
\,
\end{eqnarray}
In the first range ($\zeta \ll \Delta V^{1/3}$), we have $J_s(x)\approx J_s(\Delta V)$
and thus obtain
\begin{eqnarray*}
I_s^{(1)}(\sigma) &\approx& J_s(\Delta V)\int_{\Delta V-\zeta}^{\Delta V+\zeta} 
{dx\over\sqrt{\zeta^2-(x-\Delta V)^2}} \nonumber\\
&=& \pi J_s(\Delta V).
\end{eqnarray*}
The second range ($\zeta\gg\Delta V^{1/3}, |s-\Delta V|\ll\zeta$) similarly yields 
\begin{eqnarray*}
I_s^{(2)}(\sigma)\approx {1\over\zeta}\int_{\Delta V-\zeta}^{\Delta V+\zeta} 
J_s(x)dx \approx {1\over\zeta}.
\end{eqnarray*}
In the third range we employ the asymptotic expansion \cite{AS}
\begin{eqnarray*}
J_s(x)\sim {1\over s^{1/3}}{\rm e}^{-{2\sqrt{2}\over 3}{(s-x)^{3/2}\over\sqrt{s}}}
\end{eqnarray*}
and find
\begin{eqnarray*}
I_s^{(3)}(\sigma) &\approx& {1\over\sqrt{2\zeta}s^{1/3}}
\int_0^s {dx\over\sqrt{s-x}}\,{\rm e}^{-{2\sqrt{2}\over 3}{(s-x)^{3/2}\over\sqrt{s}}}\nonumber\\
&\approx& {\Gamma\left({1\over 3}\right)\over 3^{2/3}\sqrt{\zeta s^{1/3}}}
\sim {1\over\sqrt{\zeta\Delta V^{1/3}}}\ ,
\end{eqnarray*}
where $\Gamma$ denotes the Gamma function. This shows Eq.\,(\ref{Isigma2}).

Now we can evaluate the integral 
$I_s\equiv\int_0^\infty d\sigma \,\tilde\Phi(\sigma) I_s(\sigma)$ over $\sigma$
in the corresponding regions:
\begin{eqnarray}
\label{I1}
I_s^{(1)} &=& 
{2m^2\over|v_0|}\int_0^{\Delta V^{1/3}}d\zeta
\,\tilde\Phi(\sigma)\pi J_s(\Delta V) \nonumber\\
&\approx& 
{2m^2\over|v_0|}\Delta V^{1/3}\,\tilde\Phi(0)\pi J_s(\Delta V)\,,
\end{eqnarray}
where $\tilde\Phi(\sigma)$ was approximated by $\tilde\Phi(0)$, 
since the range of $\zeta$ values corresponds to $\sigma\ll m^2\alpha^2$.
With the abbreviation $y\equiv 2m^2a_0^2/|v_0|$, we find in the second region
\begin{eqnarray}
\label{I2}
I_s^{(2)} &=& {2m^2\over|v_0|} \int_{\Delta V^{1/3}}^\infty {d\zeta\over\zeta}
{\tilde\Phi(0)\over (1+y\zeta)^2} \nonumber\\
&=& {2m^2\over|v_0|} \tilde\Phi(0) \left[ \ln\left(1+y\Delta V^{1/3}\right) 
- \ln\left(y\Delta V^{1/3}\right)\right. \nonumber\\
& & \ \ \ \ \ \ \ \ \ \ \ \ \ \  - \left.{1\over 1+y\Delta V^{1/3}} \right] \nonumber\\
&\approx& {2m^2\over|v_0|} \tilde\Phi(0) \left[
\ln\left({\alpha^2|v_0|\over 8\Delta V^{1/3}}\right) -1\right],
\end{eqnarray}
where $y\Delta V^{1/3}=8\Delta V^{1/3}/(\alpha^2|v_0|)\ll 1$ 
was exploited in the last step.
Note that we have not restricted the range of integration according to 
$|s-\Delta V|\ll\zeta$ here, since $1/\zeta \ll 1/\sqrt{\zeta\Delta V^{1/3}}$.
For calculating the third region, we do restrict the integration to satisfy
$|s-\Delta V|\sim \zeta$ as follows:
\begin{eqnarray}
\label{I3}
I_s^{(3)} &\approx& {2m^2\over|v_0|} 
\int_{|s-\Delta V^{1/3}|-\Delta V^{1/3}}^{|s-\Delta V^{1/3}|+\Delta V^{1/3}}
{\tilde\Phi(\zeta)d\zeta\over\sqrt{\zeta\Delta V^{1/3}}} \nonumber\\
&\approx& {2\tilde\Phi(0)\over\Delta V^{1/6}}\left(
\sqrt{|s-\Delta V|+\Delta V^{1/3}} \right. \nonumber\\
& & \ \ \ \ \ \ \ \ \ \  - \left. \sqrt{|s-\Delta V|-\Delta V^{1/3}}\ \right) \nonumber\\
&\approx& 2\tilde\Phi(0)\sqrt{\Delta V^{1/3}\over |s-\Delta V|}\ ,
\end{eqnarray}
where in the last step $\Delta V^{1/3}\ll\zeta\sim|s-\Delta V|$ was applied.

Since $I_s^{(3)}\ll I_s^{(2)}\approx I_s^{(1)}$ according to Eqs.\,(\ref{I1})-(\ref{I3}), we may write $I_s\approx 2I_s^{(1)}$ and find
\begin{eqnarray*}
\bar J_{r,s} \approx 
{m^2 p_x^{\rm max}\over\pi^2}\,\tilde\Phi(0) {\Delta V^{1/3}\over|v_0|}
J_{r-2s}(U) J_s(\Delta V).
\end{eqnarray*}
By summing over $s$, we finally obtain
\begin{eqnarray*}
\bar J_r \approx \tilde\Phi(0)
{m^2\omega\over2\sqrt{2}\pi^2\xi}\left({m\xi^2\over\omega}\right)^{1/3}
{\Delta V^{1/3}\over|v_0|}J_r(U,\Delta V)\,,
\end{eqnarray*}
which coincides with Eq.\,(\ref{barJr}).

\vspace{1cm}

\renewcommand{\theequation}{C\arabic{equation}}
\setcounter{equation}{0}
\appendix
\begin{center}
  {\bf APPENDIX C}
\end{center}
In this appendix we derive the asymptotic expansion (\ref{genbesasy}) for the generalized Bessel function $J_N(U,V)$ (see also Refs.\,\cite{SFA3,Leubner,genbes2}). We consider the integral $\mathcal{I} \equiv \int dy_-\,{\rm e}^{ {\rm i}f(y_-)}$ of Eq.\,(\ref{I}).
Due to periodicity and the unbounded range of integration, Eq.\,(\ref{cosy}) defines
an infinite number of saddle points, which can be divided into four classes ($j=1,2,3,4$) according to 
\begin{eqnarray}
\omega y_-^{(j,N)} = \omega y_-^{(j)} + 2\pi N\,.
\end{eqnarray}
Here, $N$ is an integer number and the four particular solutions
\begin{eqnarray}
\label{SPsol}
\omega y_-^{(1)} &=& \arccos\left(-{U\over 8V}+\rho\right)\,, \nonumber\\
\omega y_-^{(2)} &=& 2\pi - \arccos\left(-{U\over 8V}+\rho\right)\,, \nonumber\\
\omega y_-^{(3)} &=& \arccos\left(-{U\over 8V}-\rho\right)\,, \nonumber\\
\omega y_-^{(4)} &=& 2\pi - \arccos\left(-{U\over 8V}-\rho\right)\,,
\end{eqnarray}
with
\begin{eqnarray*}
\rho \equiv \sqrt{{U^2\over 64V^2}+{\Delta Q\over 4V\omega}+{1\over 2}}\,,
\end{eqnarray*}
lie in the intervall $[0,2\pi)$. At the saddle points, the second derivative of the phase (\ref{f}) becomes
\begin{eqnarray}
\label{d2f}
{d^2f\over dy_-^2}=\pm 8\omega^2V\rho\sin(\omega y_-^{(j,N)}),
\end{eqnarray}
with the positive (negative) sign for $j=1,2$ ($j=3,4$). Using the general identity
\begin{eqnarray*}
\sum_{N=-\infty}^\infty {\rm e}^{-2\pi {\rm i} N\frac{\Delta Q}{\omega}}
&=& \omega\sum_{N=-\infty}^\infty \delta(\Delta Q-N\omega),
\end{eqnarray*}
we can sum up all contributions from the saddle points $y_-^{(1,N)}$ in closed form and obtain
\begin{eqnarray*}
\mathcal{I}^{(1)} &=& \sqrt{2\pi}\sum_{N=-\infty}^\infty \delta(\Delta Q-N\omega) \nonumber\\
&\times&\!\!\! {{\rm e}^{{\rm i}N\omega y_-^{(1)} - {\rm i}U\sin(\omega y_-^{(1)}) - 
{\rm i}V\sin(2\omega y_-^{(1)})+\sigma^{(1)}{\rm i}{\pi\over 4}}\over
 \sqrt{8\rho|V\sin(\omega y_-^{(1)})|}}\,,
\end{eqnarray*}
where $\sigma^{(1)}\equiv {\rm sgn}[V\sin(\omega y_-^{(1)})]$ fixes the sign of the
second derivative in Eq.\,(\ref{d2f}).
Analogously, we find $\mathcal{I}^{(2)}={\mathcal{I}^{(1)}}^*$ so that
\begin{eqnarray*}
\!\!\!\!\!\! & & \mathcal{I}^{(1)}+\mathcal{I}^{(2)} = 2\sqrt{2\pi}\sum_{N=-\infty}^\infty \delta(\Delta Q-N\omega) \nonumber\\
\!\!\!\!\!\! & & \times {\rm Re}\left(
 {{\rm e}^{{\rm i}N\omega y_-^{(1)} - {\rm i}U\sin(\omega y_-^{(1)}) - 
{\rm i}V\sin(2\omega y_-^{(1)})+\sigma^{(1)}{\rm i}{\pi\over 4}}\over
 \sqrt{8\rho|V\sin(\omega y_-^{(1)})|}} \right).
\end{eqnarray*}
Note that $\sin(\omega y_-^{(2)})=-\sin(\omega y_-^{(1)})$.
The contributions of the saddle points $y_-^{(3,N)}$ and $y_-^{(4,N)}$ add up
similarly and we finally obtain
\begin{widetext}
\begin{eqnarray}
\label{Ifin}
\mathcal{I} &=& \mathcal{I}^{(1)}+\mathcal{I}^{(2)}+\mathcal{I}^{(3)}+\mathcal{I}^{(4)} 
   = 2\sqrt{2\pi}\sum_{N=-\infty}^\infty \delta(\Delta Q-N\omega)\nonumber\\
& & \times\ {\rm Re}\left(
 {{\rm e}^{{\rm i}N\omega y_-^{(1)} - {\rm i}U\sin(\omega y_-^{(1)}) - {\rm i}V\sin (2\omega y_-^{(1)}) +\sigma^{(1)} {\rm i}{\pi\over 4}}\over
 \sqrt{8\rho|V\sin(\omega y_-^{(1)})|}} +
{{\rm e}^{{\rm i}N\omega y_-^{(3)} - {\rm i}U\sin(\omega y_-^{(3)}) - {\rm i}V\sin (2\omega y_-^{(3)}) -\sigma^{(3)} {\rm i}{\pi\over 4}}\over
 \sqrt{8\rho|V\sin(\omega y_-^{(3)})|}} \right) 
\end{eqnarray}
\end{widetext}
with $\sigma^{(3)}\equiv {\rm sgn}[V\sin(\omega y_-^{(3)})]$.
A comparison of Eq.\,(\ref{Ifin}) with the last line of Eq.\,(\ref{Ialtern}) confirms the asymptotic representation (\ref{genbesasy}) for the generalized Bessel function.

We comment on the applicability range of Eq.\,(\ref{genbesasy}) \cite{Leubner}. First of all, in the spirit of the saddle-point method, large values $|U|, |V|, |N|\gg 1$ are required. The saddle points (\ref{SPsol}) should be real and well separated along the real axis. This implies $\rho^2>0$ and $|\cos(\omega y_-)|< 1$. If not all saddle points are real, then the contribution from the complex ones may be ignored to a good approximation, as illustrated below (cf. also Eq.\,(3.7) in \cite{Leubner}). Moreover, the value of the second derivative $d^2f/dy_-^2$ in Eq.\,(\ref{d2f}) must not be too close to zero; otherwise the next-order term in the Taylor expansion of the exponent (\ref{f}) becomes dominant. A comparison of the asymptotic formula (\ref{genbesasy}) with the exact evaluation of $J_N(U,V)$ via the series expansion (\ref{genbes}) is displayed in Fig.\,4. We emphasize that only real saddle points have been taken into account here; contributions from saddle points with non-zero imaginary part are exponentially damped. I.e., in this case only one of the two terms within the real-part function in Eq.\,(\ref{genbesasy}) was considered. We see that the asymptotic expansion provides a very good approximation, in general. Only in a small region around $U=15$, $V=-15$ in Fig.\,4(a) [or $U=80$, $V=-15$ in Fig.\,4(c)] the asymptotic formula fails. As shown in Fig.\,4(b), it is exactly at this point where $d^2f/dy_-^2$ vanishes. In order for Eq.\,(\ref{genbesasy}) to be applicable, the parameters need to fulfill the condition
\begin{eqnarray}
\label{f2f3}
\sqrt{\frac{2 |f''|}{\pi}} > \frac{3}{\Gamma\left(\frac{1}{3}\right)}\sqrt[3]{\frac{|f'''|}{6}}.
\end{eqnarray}
Here, $f''=\pm 8V\rho\sin(\omega y_-)$ and $f'''=8V[\pm\rho\cos(\omega y_-)-\sin^2(\omega y_-)]$ denote the second and third derivatives of $f$ with respect to $\omega y_-$, taken at the potential saddle-point positions $\omega y_-^{(j)}$ according to Eq.\,(\ref{SPsol}). When $f''$ tends to zero, the condition (\ref{f2f3}) will be violated eventually, since $f'''$ remains finite [see Fig.\,4(b)]. In this situation, moreover, $|\cos(\omega y_-)|$ tends to unity, so that the two respective saddle points approach each other. Then the isolation criterion is no longer satisfied.

\begin{figure}[ht]
\begin{center}
\includegraphics[width=7cm]{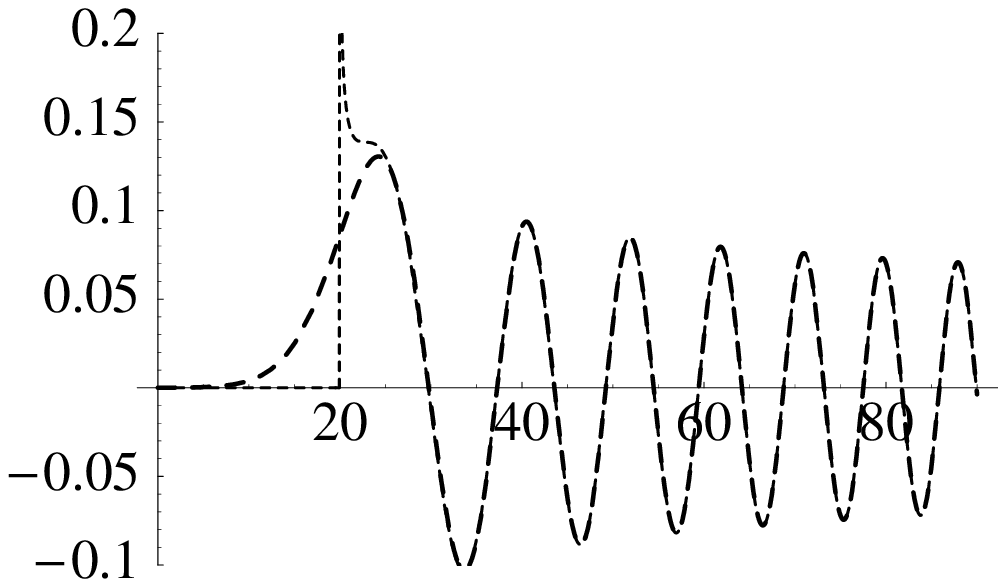}
\begin{picture}(0,0)(0,0)
\put(-215,110){\large (a)}
\put(0,40){\large$U$}
\put(-70,85){\large$J_{50}(U,15)$}
\end{picture}
\includegraphics[width=7cm]{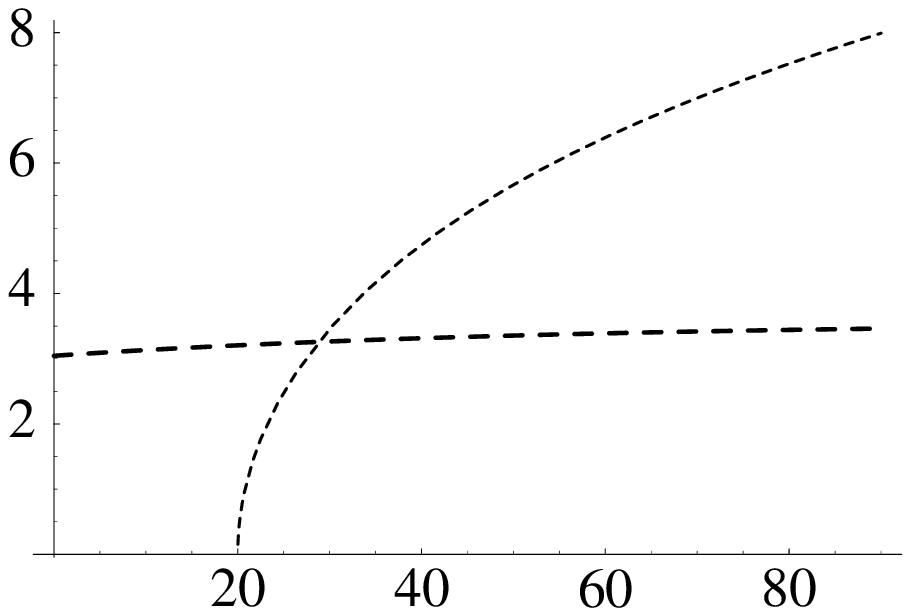}
\begin{picture}(0,0)(0,0)
\put(-215,110){\large (b)}
\put(-192,10){\large$0$}
\put(-185,4){\large$0$ }
\put(-8,10){\large$U$}
\put(-110,100){\large$\sim \sqrt{f''}$}
\put(-70,45){\large$\sim \sqrt[3]{f'''}$}
\end{picture}
\includegraphics[width=7cm]{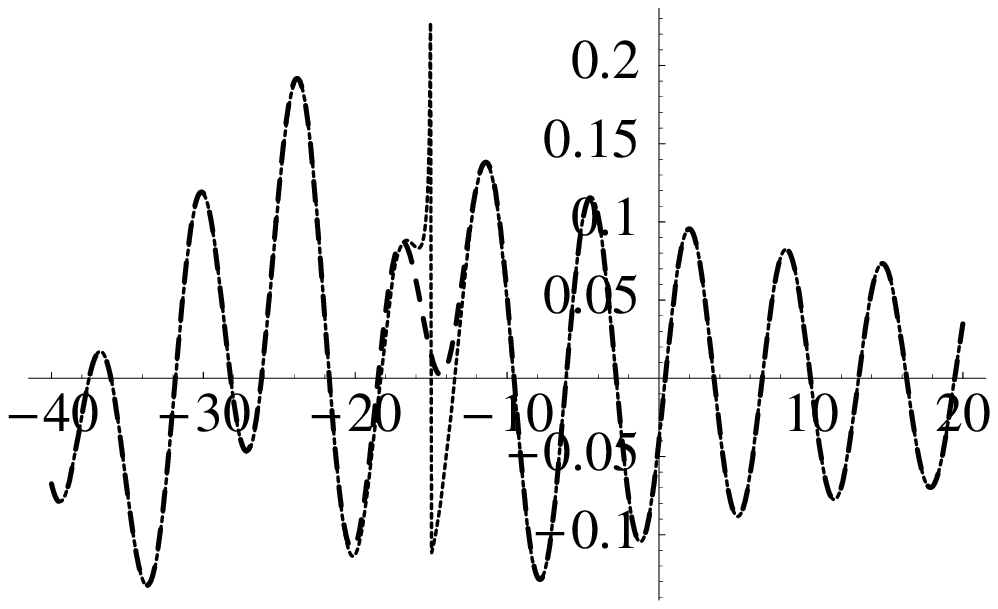}
\begin{picture}(0,0)(0,0)
\put(-215,110){\large (c)}
\put(-2,45){\large$V$}
\put(-55,105){\large$J_{50}(80,V)$}
\end{picture}
\caption{\label{GenBes} Illustration of the applicability range of the asymptotic expansion in Eq.\,(\ref{genbesasy}) for the generalized Bessel function. (a) $J_{50}(U,15)$ in the range $0\le U\le 90$; the thick dashed line shows the exact value resulting from the series expansion in Eq.\,(\ref{genbes}), the thin dotted line shows the asymptotic formula. (b) The condition (\ref{f2f3}) for the example of the figure in (a), with the thick dashed and thin dotted lines showing the left- and right-hand sides of Eq.\,(\ref{f2f3}), respectively. (c) Same as (a), but for $J_{50}(80,V)$ in the range $-40\le V\le 20$.}
\end{center} 
\end{figure}

In our numerical calculations the asymptotic expansion (\ref{genbesasy}) is applied to evaluate $J_r(U,\Delta V)$ in Eq.\,(\ref{barJr}). We check that the above conditions of applicability are satisfied in the integration region of interest. In accordance with the kinematical analysis in Sec.\,III.A, the characteristic magnitudes of the parameters are $r\approx |U|\sim m\xi^2/\omega\sim 10^{10}$ and $|\Delta V|\sim |V|/10\approx m\xi^2/(20\omega)\sim 10^9$. The parameter $\rho$ is therefore positive and can be approximated by a Taylor expansion as
\begin{eqnarray*}
\rho &\approx& \frac{|U|}{8|\Delta V|}\left( 1 + \frac{8r\Delta V}{U^2} + \frac{16\Delta V^2}{U^2} \right) \\
&=& \frac{|U|}{8|\Delta V|} \pm \frac{r}{|U|} + \frac{2|\Delta V|}{|U|}\,,
\end{eqnarray*}
where the first term typically is larger than unity, the second term is close to unity, and the last term is much smaller than unity.
According to Eq.\,(\ref{cosy}), there are real saddle points $z$ located at
\begin{eqnarray}
\label{cosz}
|\cos(\omega z)| \approx \frac{r \pm 2|\Delta V|}{|U|} 
\approx  1 - \frac{2|\Delta V|}{|U|} \lesssim 1.
\end{eqnarray}
From a physical point of view we note that a cosine value close to unity agrees with the discussion in Sec.\,III.B, since the electron-positron annihilation occurs close to the field maxima (i.e., at $|\cos(\omega x_-)|\approx 1$) and the muons are created shortly afterwards. From Eq.\,(\ref{cosz}), we obtain $|\sin(\omega z)|\approx 2\sqrt{|\Delta V/U|}\lesssim 1$. This implies $|f''|\sim |f'''|\sim 8|\Delta V|\gg 1$ and the condition (\ref{f2f3}) is satisfied. 
As a result, the asymptotic expansion (\ref{genbesasy}) is applicable for our purposes of numerical computation in the mainly contributing regions of the integration domain in Eq.\,(\ref{RPs}).


\begin{thebibliography}{99}
\bibitem{stable} On the typical time scale of strong laser pulses ($\tau \sim$ fs$-$ps), a free Ps atom can be considered as stable since its lifetime is $\sim 10^{-7}$\,s (ortho-Ps) or $\sim 10^{-10}$\,s (para-Ps), respectively. In a statistical mixture of Ps atoms, the ortho-triplett has a relative abundance of 75\%.
\bibitem{resonances} For certain values of $p_\pm$ and $n$ the denominator $q^2=(q_++q_--nk)^2$ in Eq.\,(\ref{See2}) becomes zero, which indicates a resonance behaviour of the amplitude (see, e.g., Refs. \cite{Moller,Bhabha,electron-muon}). One can show, however, that these resonances give a negligible contribution to the total process probability in the present situation \cite{muonPRD}.
\end{thebibliography}
\end{document}